\documentclass[final,5p,times,twocolumn,authoryear]{elsarticle}

\usepackage{amssymb}
\usepackage{amsmath}

\journal{Computer Networks}

\usepackage[utf8]{inputenc}
\usepackage{multicol}
\usepackage{hyperref}
\usepackage{float}
\usepackage{amsmath}          
\usepackage{amssymb}          
\usepackage{amsthm}
\usepackage{comment}
\usepackage{doi}
\usepackage{float}
\usepackage{graphicx}
\usepackage{parskip}
\usepackage{url}
\usepackage{siunitx}
\usepackage{mathtools}
\usepackage{tikz}
\usetikzlibrary{positioning}
\usetikzlibrary{decorations.text}
\usetikzlibrary{decorations.pathmorphing}
\usepackage{enumitem}
\usepackage{pdflscape}
\usepackage{caption, booktabs}
\usepackage{balance}
\usepackage{scalerel}
\usepackage{tikz}
\usepackage{longtable}
\usepackage{multirow}
\usepackage{tabularx}
\usetikzlibrary{svg.path}
\usepackage{svg}
\usepackage{subcaption}
\usepackage{orcidlink}

\begin{document}

\begin{frontmatter}



\title{Impact of Cross Technology Interference on Time Synchronization and Join Time in Low-Power Wireless Networks}


\author[first]{\orcidlink{0009-0009-5141-8160} Zegeye Mekasha Kidane \corref{cor1}}
\affiliation[first]{organization={Max planck Institute for Radio Astronomy},           
            addressline={Electronics Division}, 
            city={Bad Muenstereifel},
            postcode={53902}, 
            state={NRW},
            country={Germany}            
            }
        \ead{zkidane@mpifr.de}
        \cortext[cor1]{corresponding author.}
\author[second]{\orcidlink{0000-0002-7911-8081} Waltenegus Dargie}
\affiliation[second]{organization={Technische Universitaet Dresden},            
            addressline={Faculty of Computer Science}, 
            city={Dresden},
            postcode={01062}, 
            state={Sachsen},
            country={Germany},            
            }
        \ead{waltenegus.dargie@tu-dresden.de}
         \corref{corresponding author}
\begin{abstract}   
Low-power and low-cost wireless sensor networks enable scalable and affordable sensing and can be deployed in different environments to monitor various physical parameters. In some environments, these networks may have to coexist and interact with other systems which use the same frequency spectrum for communication. This potentially results in cross-technology interference (CTI). Dynamic channel hopping is one of the mechanisms that is currently employed to deal with CTI, but its usefulness depends on the channel selection and occupation timing. In this paper, we experimentally study the impact of CTI (caused by IEEE 802.11 networks) on time synchronization and network join time. Experiment results show that CTI can increase time drift between a child and a parent node by up to $\pm 3$ clock ticks between two synchronization intervals. Likewise, CTI affects new nodes from timely joining a network. In a simple network which does not involve multi-hop communication, the time it takes for nodes to join the network in the absence of CTI is between 40 and 70 ms (83.3\% of the time). In the presence of CTI, 96.82\% of the time, the join time is between 100 and 200 ms. In other words, the join time in the presence of CTI is about five times higher. Interestingly, not only the main spectral lobes, but also the spectral sidelobes of interfering networks impact the performance of low-power networks. 
\end{abstract}




\end{frontmatter}

\title{
Impact of Cross Technology Interference on Time Synchronization and Join Time in Low-Power Wireless Networks
}

\begin{abstract}
\begin{abstract}   
Low-power and low-cost wireless sensor networks enable scalable and affordable sensing and can be deployed in different environments to monitor various physical parameters. In some environments, these networks may have to coexist and interact with other systems which use the same frequency spectrum for communication. This potentially results in cross-technology interference (CTI). Dynamic channel hopping is one of the mechanisms that is currently employed to deal with CTI, but its usefulness depends on the channel selection and occupation timing. In this paper, we experimentally study the impact of CTI (caused by IEEE 802.11 networks) on time synchronization and network join time. Experiment results show that CTI can increase time drift between a child and a parent node by up to $\pm 3$ clock ticks between two synchronization intervals. Likewise, CTI affects new nodes from timely joining a network. In a simple network which does not involve multi-hop communication, the time it takes for nodes to join the network in the absence of CTI is between 40 and 70 ms (83.3\% of the time). In the presence of CTI, 96.82\% of the time, the join time is between 100 and 200 ms. In other words, the join time in the presence of CTI is about five times higher. Interestingly, not only the main spectral lobes, but also the spectral sidelobes of interfering networks impact the performance of low-power networks. 
\end{abstract}
\end{abstract}



\section{Introduction}
\label{sec:intro}

Cross technology interference (CTI) arises as a consequence of two or more heterogeneous wireless technologies utilizing the same radio spectrum seize the medium simultaneously. These systems do not have a shared medium access control (MAC) protocol to arbitrate between them. Often, the impact of CTI is asymmetric, as the systems rely on different physical layer strategies, the ones using higher transmission powers and more advanced modulation schemes affecting those with lower transmission powers and less advanced modulation schemes more significantly \cite{dargie2024mitigating}. In the context of low-power wireless sensing networks, the most important CTI arises from IEEE 802.11 networks, since both systems rely on the 2.4 GHz, license-free, spectrum \cite{shi2017signal}. 

Different approaches have been proposed to mitigate CTI in this spectrum. Some of these enable low-power networks to estimate the transmission and channel occupation pattern of the more advanced networks to determine least affected channels and best transmission timing \cite{hithnawi2016crosszig}. Other approaches aim to establish back-channels for interfering networks to  convey transmission intentions and coordinate packet transmission \cite{yin2018explicit, guo2020zigfi}. But one of the most widely used and practical solutions is dynamic channel/frequency hopping \cite{hermeto2017scheduling}. Like many other technologies, such as the IEEE 802.15.1 technologies  \cite{golmie2003bluetooth}, dynamic channel hopping entails dynamic channel selection, which requires exact timing and coordination between communicating partners, so that both the  transmitter and the receiver transit from a present channel to a future channel in sync \cite{chang2015adaptive, duquennoy2017tsch}. This feature makes the very solution which deals with CTI, vulnerable to it, since in the presence of CTI, time synchronization becomes a serious challenge. In this paper we investigate this challenge experimentally. As the first contribution of this paper, both through field deployments and experiments in controlled environments, we investigate the impact of CTI on the link quality of low power networks, closely examining the impact in all the available channels for the case of IEEE 802.15.4 networks. As a second contribution, we closely examine the  impact of CTI on the synchronization and network join latency, assuming that the IEEE 802.15.4 networks employ the Time-Slotted and Channel Hopping (TSCH) medium access protocol \cite{tinka2010decentralized}. 

The rest of this paper is organised as follows: In Section \ref{sec: related}, we review related work. In Section \ref{sec:background}, we provide background information and discuss field deployments and experiments setups. In Section \ref{sec:lq} we address the impact of CTI On link quality in low-power networks. In Section \ref{sec:ts}, we examine the impact of CTI on time synchronization, and in  Section \ref{sec:latency}, on self-organization. Finally, in \ref{sec:conclusion}, we provide concluding remarks and outline future work.

\section{ Related Works} 
\label{sec: related}

Time synchronization is a critical component for a large number of network services and applications. Physical and MAC layer protocols often rely on accurate timing to efficiently utilise a shared medium, effectively communicate, and maximize performance. Similarly, higher-level applications, such as localization and tracking applications, require time synchronization to deliver reliable services. CTI inhibits nodes from synchronizing clocks and, this, in turn, not only leads to a deterioration of quality of service, but also further exacerbate interference.

In \cite{Reitz_FMCW_interference_2024}, the authors investigate the impact of device imperfection (slight variations in chirp forms, attributable to different hardware properties such as clock drift) on mutual interference in radar sensors. The authors remark that one of the most critical challenges in the mitigation of this type of interference is the lack of synchronization in terms of both time and frequency. This challenge, compounded by variations in operational parameters (frequency slope, up, down, inter-chirp duration, sweep bandwidth, and Analog-to-Digital Converter (ADC) sampling time), leads to a complex interference behaviour. Specifically, their investigation reveals that the impact of parallel chirp interference, as minor deviations in the chirp slope, leads to a significant Intermediate Frequency (IF) interference, characterized by considerable interference power. In \cite{guo_imperfect_synchronization_2020}, the authors investigate the impacts of imperfect synchronization and channel estimation on known interference cancellation. The authors propose two different models to normalize time and frequency synchronization errors. Accordingly, the normalized time synchronization error is modelled as inter-symbol interference; and frequency and phase errors, as inter-frequency interference. The known interference rejection ratio is derived for the multipath block fading channel by jointly considering incomplete time-frequency synchronization and channel estimation as well as phase noise.

The work in \cite{Luz_lorawan_block_interference_2020} focuses on Long Range (LoRa) networks. The authors experimentally determined the extent to which a LoRa channel is vulnerable to jamming by an electromagnetic interference, and propose a model to estimate an immunity region. The study is closer to ours in that it, too, statistically analyzes link quality metrics (packet loss, RSSI, and SNR) to characterize interference and network performance. Experimental results show that RSSI values during interference are typically higher than values without interference. Our experiment results agree with theirs. In \cite{Tan_uav_tsynchronization_2020}, the authors propose a UAV-assisted, low-power consumption, time synchronization algorithm based on cross-technology communication (CTC) \cite{kim2015freebee} for a large-scale wireless sensor network. The algorithm enables the UAV to employ a high-power to broadcast time synchronization packet to the ground nodes. The implementation and field experiment consisted of 30 low-power RF-CC2430 nodes and a DJI M100 UAV on a 1 km highway and an indoor site. The experiment results suggest  that time synchronization could be achieved with a synchronization error below $30$ us. This result agrees with our experimental result for the case of a modest CTI. 

In \cite{Gao_tsynch_cti_2023}, the authors investigate the problem of time synchronization among coexisting heterogeneous technologies such as WiFi, ZigBee, and Bluetooth. The authors propose a time synchronization strategy which relies on CTC and enables WiFi devices to assist ZigBee devices during time synchronization. Accordingly, a nearby WiFi device acts as a coordinator node and broadcasts timestamps to the ZigBee devices. The ZigBee nodes update their clocks periodically depending on these timestamps. In addition, the WiFi and ZigBee devices negotiate for spectrum allocation to reduce the impact of CTI. When receiving a timestamp from a WiFi coordinator, a low-power node enters into clock calibration; adjusts its local clock accordingly; and handles interference through negotiation. Both analytic and experiment results show that the proposed approach achieves a global time synchronization with a time error lower than $50$ us.

In \cite{Salazar_Lopez_gps_synchronization_2024}, the authors propose a combination of Pulse Per Second (PPS) synchronization with the help of a GPS  and real-time clocks (RTC) for structural health monitoring. The PPS signal provided by a GPS is used as a trigger signal for data synchronization. The authors observe that PPS, alone, however, is not sufficient, as there can be nodes in the network which achieve   exact synchronization and others which are not synchronized due to various reasons (data corruption and other environmental factors), in which case, peer-to-peer synchronization using RTC is employed to achieve network-wide synchronization. 

\section{Background}
\label{sec:background}

CTI affecting low-power sensing networks has been studied mainly with respect to interference arising from IEEE 802.11 (WiFi) and IEEE 802.15.1 (Bluetooth) technologies \cite{chi2019concurrent, li2017webee}. The type of networks these technologies typically establish require transmission ranges in the order of a few hundred meters. Such networks can be found in residential, industrial, and office environments \cite{grimaldi2020autonomous, elias2014cross}. More recently, autonomous systems complying with IEEE 802.11 standards and requiring longer transmission ranges have been populating the market. These include Unmanned Aerial Vehicles (UAV) and Unmanned Surface Vessels (USV). These systems rely on wireless links to communicate with their remote control stations as well as their peers. Their transmission range far exceeds that of conventional WiFi networks. Moreover, they have stringent safety requirements, which necessitate the use of a high transmission power and reliable communication.

Interestingly, these systems can be jointly deployed with wireless sensor networks to monitor remote and vast areas. For example, in water quality monitoring, the sensor networks can be deployed on the surface of restless waters; and the UAVs and the USVs can be deployed to configure the sensor networks and to collect data from them. Therefore, investigating the effect of CTI on the coexistence and collaboration of such systems is important.  

\begin{figure}[t!]
	\centering
	\includegraphics[width=0.45\textwidth]{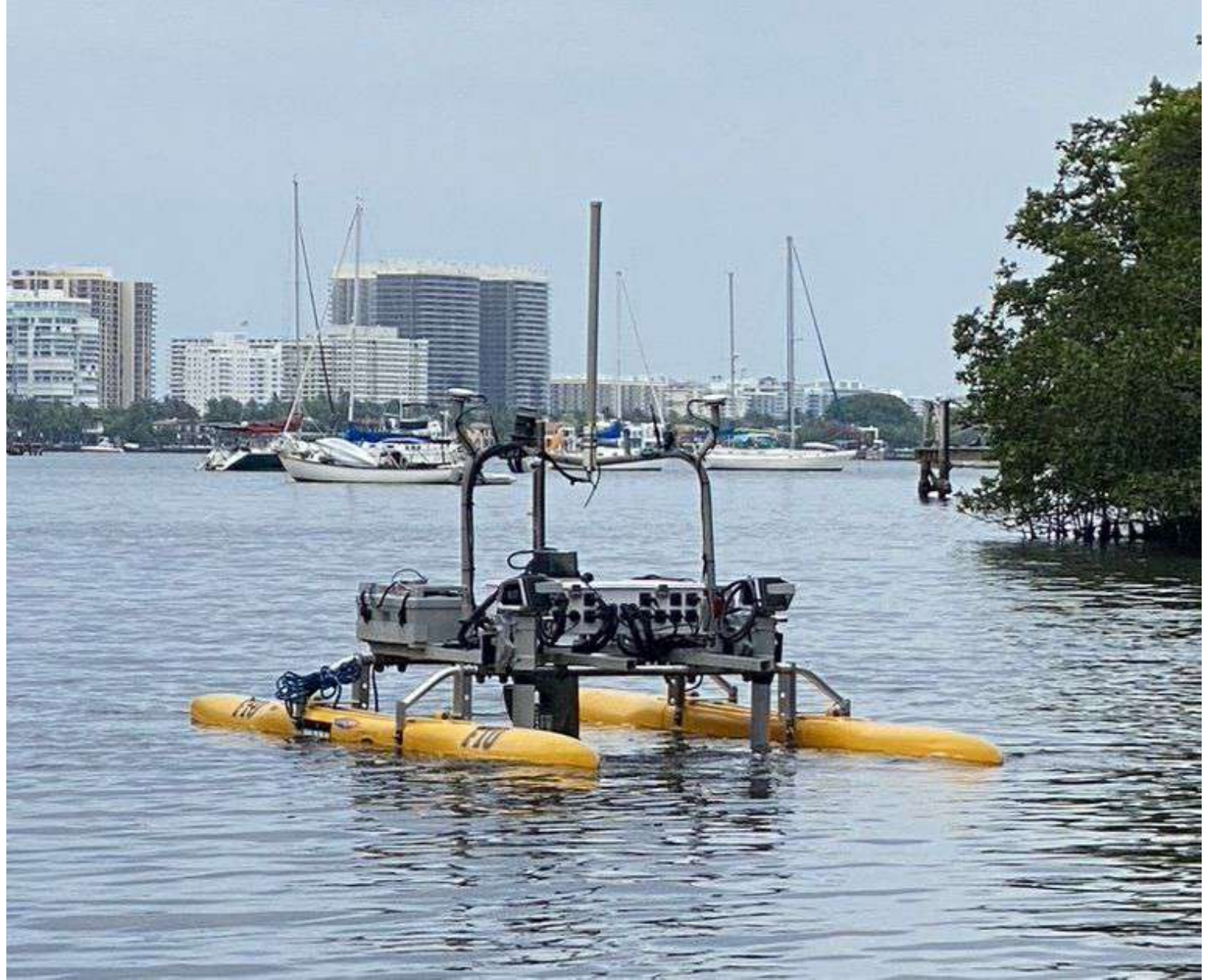}
	\caption{A USV producing cross technology interference. Deployment: North Biscayne Bay, South Florida.}
	\label{fig:mmu}
\end{figure}

\subsection{Field Deployments}

In order to have a first hand experience of the impact of CTI, we carried out deployments involving USVs and wireless sensor networks. The first deployment took place in North Biscayne Bay, Miami, Florida. The deployment consisted of a USV\footnote{A product of SeaRobotics Corporation, which used the IEEE 802.11/b for communicating with its remote station: https://www.searobotics.com/} and a network of 6 wireless sensor nodes. The nodes established a multi-hop network using a 2.4 GHz radio (CC2538). The distance between them was about 50 m. When the USV was not around, the nodes communicated with one another with a modest amount of packet loss (less than 10\%). When, the USV was within 300 m radius or so, communication was considerably constrained; now packet loss varying from 30 to 100\%. We tested all the 16 available IEEE 802.15.4 channels to avoid CTI, but performance remained poor. 

\begin{figure}[t!]
	\centering
	\includegraphics[width=0.45\textwidth]{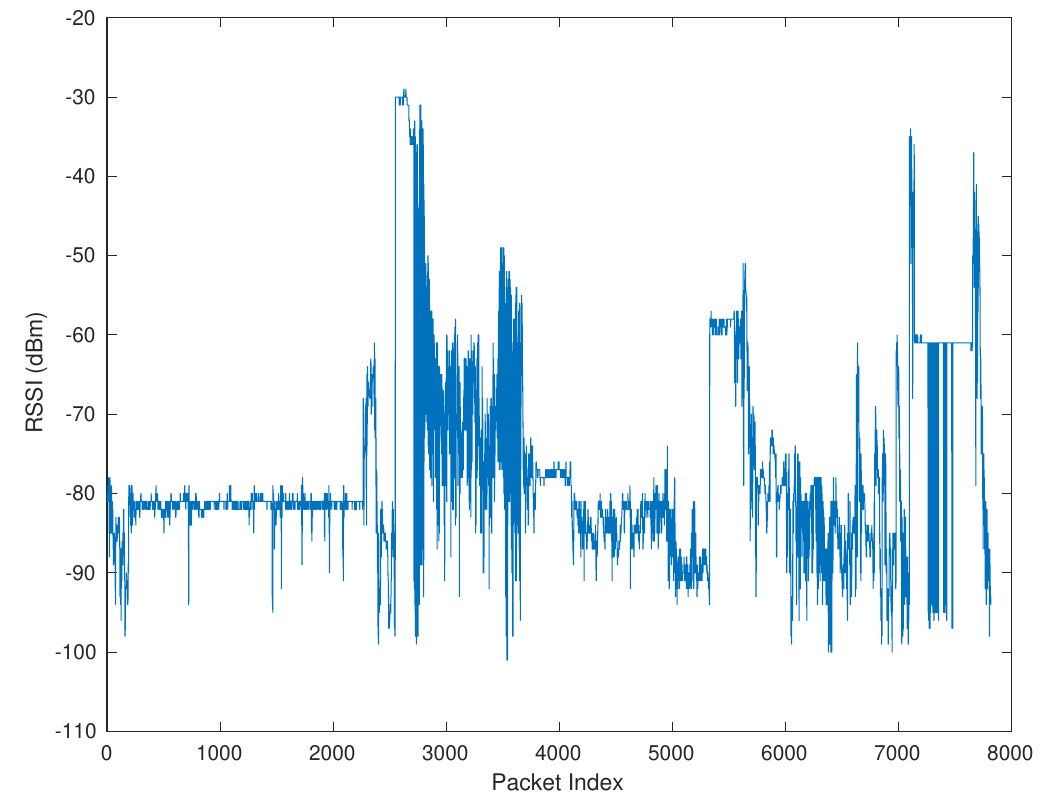}
	\caption{Link quality fluctuation in the presence of CTI (Deployment on a lake on FIU's Main Campus).}
	\label{fig:cti_fiu}
\end{figure}

The second deployment took place at one of the lakes on Florida International University's Main Campus. This time, we deployed 5 wireless sensor nodes on the surface of the lake and an additional sensor node on the USV itself. Both this node and the nodes deployed on the surface of the lake communicated with a gateway node placed outside the lake (ref. to Fig.~\ref{fig:mmu}). The present USV had a more complex setup than the one deployed in North Biscayne Bay. It communicated with its remote control station using a proprietary transceiver, operating in the 4.9-5.8 GHz band, but in addition, the control station was remotely controlled by a human agent using the IEEE 802.11b standard. When the USV navigated autonomously, both the node deployed on the boat and on the lake experienced no interference and the link quality was stable; as soon as a human agent interacted with the boat using the IEEE 802.11b interface, all the nodes experienced a significant CTI. The node which was affected the most was the one carried by the autonomous boat. Fig.~\ref{fig:cti_fiu} shows the link quality of this node, as reflected by the RSSI of the packets it received from  the base station.    

\begin{figure}[t!]
	\centering
	\includegraphics[width=0.45\textwidth]{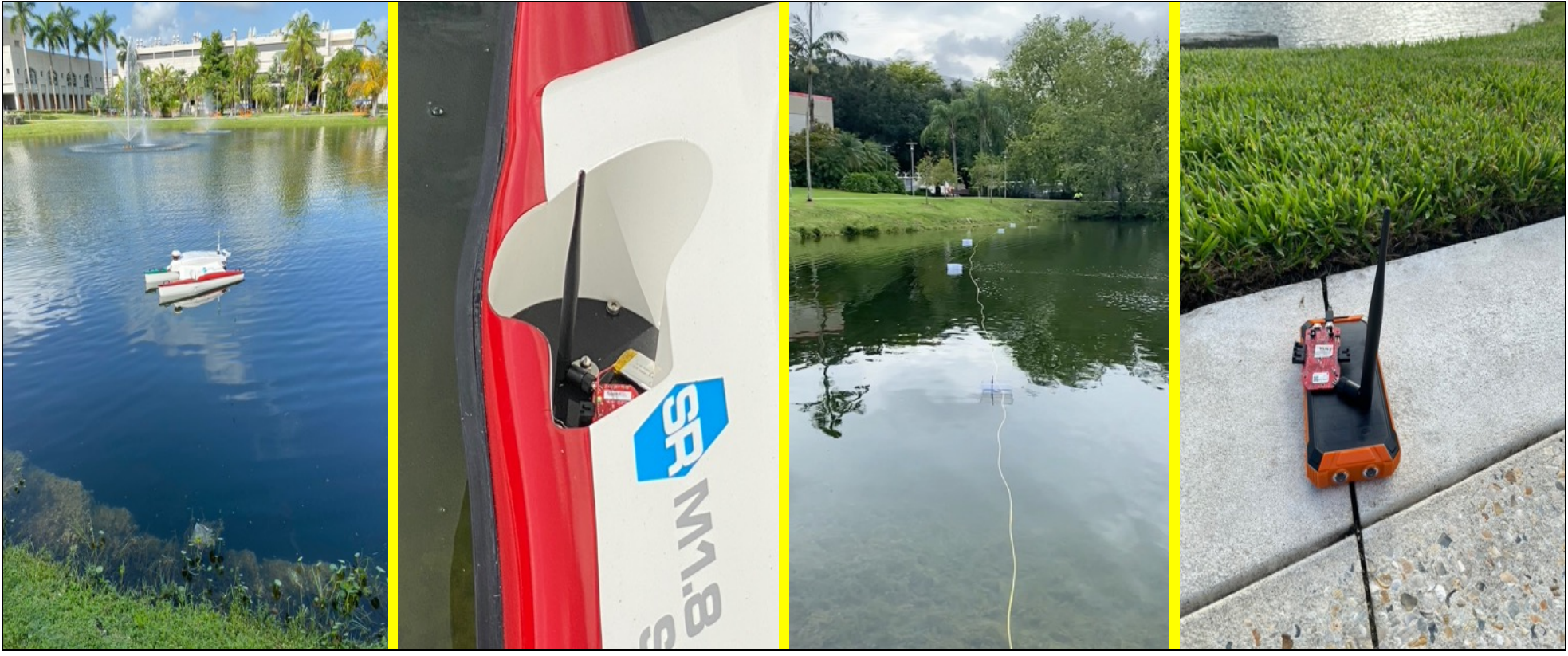}
	\caption{Deployment of a USV and a Wireless Sensor Network at one of the Lakes on Florida International University main campus. }
	\label{fig:mmu}
\end{figure}

In order to separate the effect of CTI from the effect of the motion of water on the link quality of the low-power networks, we repeated the field experiments in a controlled environment. In this setup, the low-power network consisted of three wireless sensor nodes, one of which served as a coordinator. In addition, we had two laptops and three smartphones. One of the laptops served as a programming and configuration platform as well as a gateway to the Internet. It ran \texttt{pySerial}\footnote{\begin{url}https://pyserial.readthedocs.io/en/latest/pyserial.html\end{url}.}, for executing python scripts and facilitating a serial communication with the coordinator. We established a database management system using \texttt{phpMyAdmin}\footnote{https://www.phpmyadmin.net/} and \texttt{MySQL} to export and store link quality metrics. The second laptop and the three smartphones were used to generate and induce interference.

In both types of experiments, the sensor nodes integrated the CC2538 system-on-chip \cite{cc2538_soc_2013}. The radio on this chip complies  with the IEEE 802.15.4 specification. It has 16 available channels (numbered from 11 to 26); each channel has a bandwidth of 2 MHz and a 5 MHz guard band separates adjacent channels. The radio uses the 2.4 GHz license-free spectrum; it has an adjustable transmission power (the maximum transmission power being 7 dBm), and a transmission rate of 250 kbps. The Contiki operating system \cite{Contiki-NG-2022} was used as the operating system to manage the sensor nodes and the protocols they ran.

\begin{figure}
	\centering
	\includegraphics[width=0.45\textwidth]{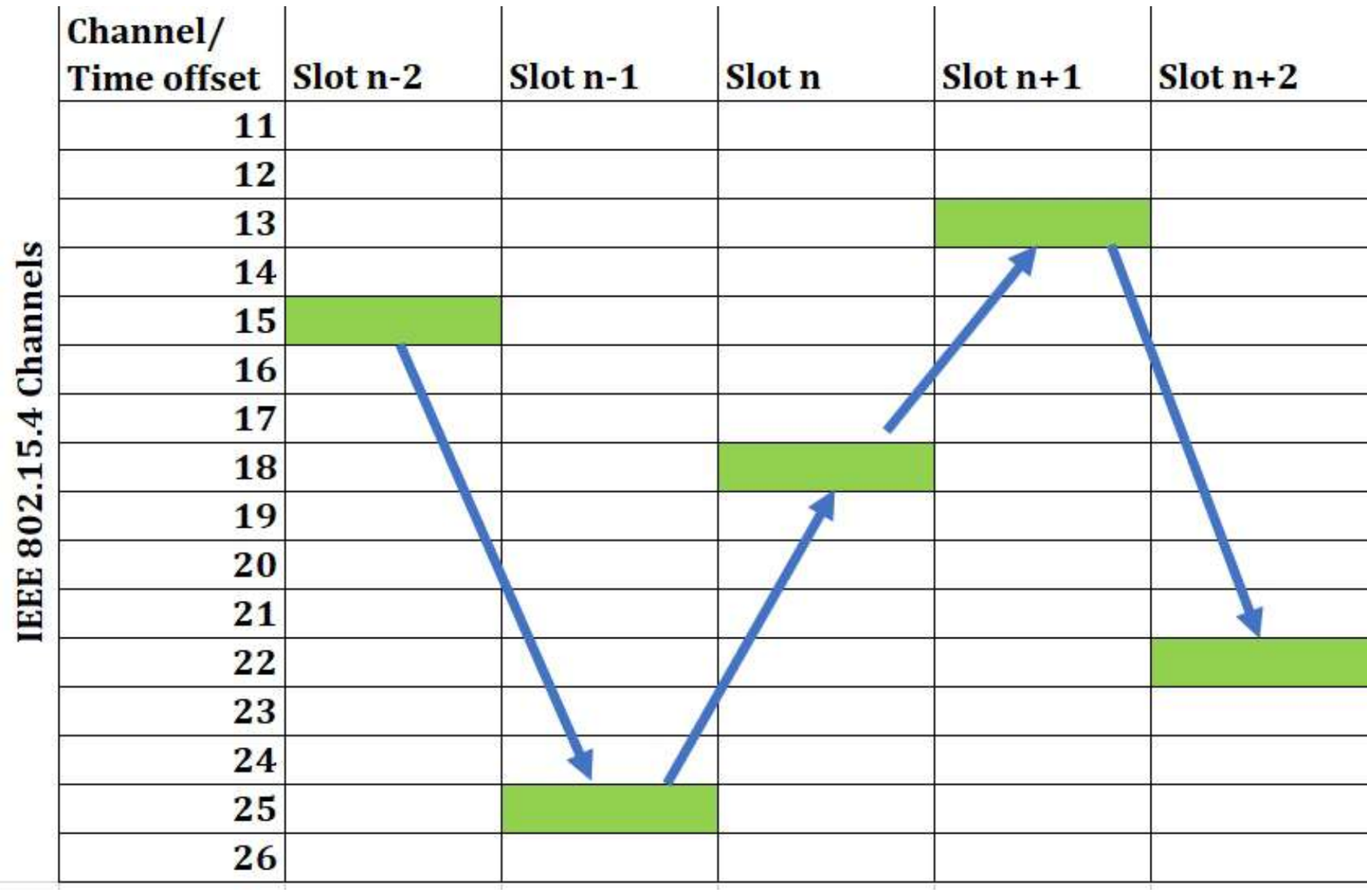}
	\caption{Illustration of slot and channel assignment in TSCH.}
	\label{fig:tsch}
\end{figure}

\subsection{Channel Assignment}

The 3 non-overlapping channels mostly used by IEEE 802.11 networks are Channels 1, 6, and 11. These channels overlap with some of the IEEE 802.15.4 channels. Channel 1 overlaps with Channels 11 to 14; Channel 6, with Channels 16 to 19; and Channel 11, with Channels 21 to 24. The three non-overlapping IEEE 802.11 channels are rarely used simultaneously, in which case, the IEEE 802.15.4 networks can employ dynamic channel selection/channel-hopping to minimize CTI. Furthermore, Channels 15, 20, 25, and 26 of the IEEE 802.15.4 networks are the least affected by IEEE 802.11 channel assignment and this knowledge can be useful during dynamic channel selection.

\begin{figure*}[t!]
	\centering
	\includegraphics[width=0.8\textwidth]{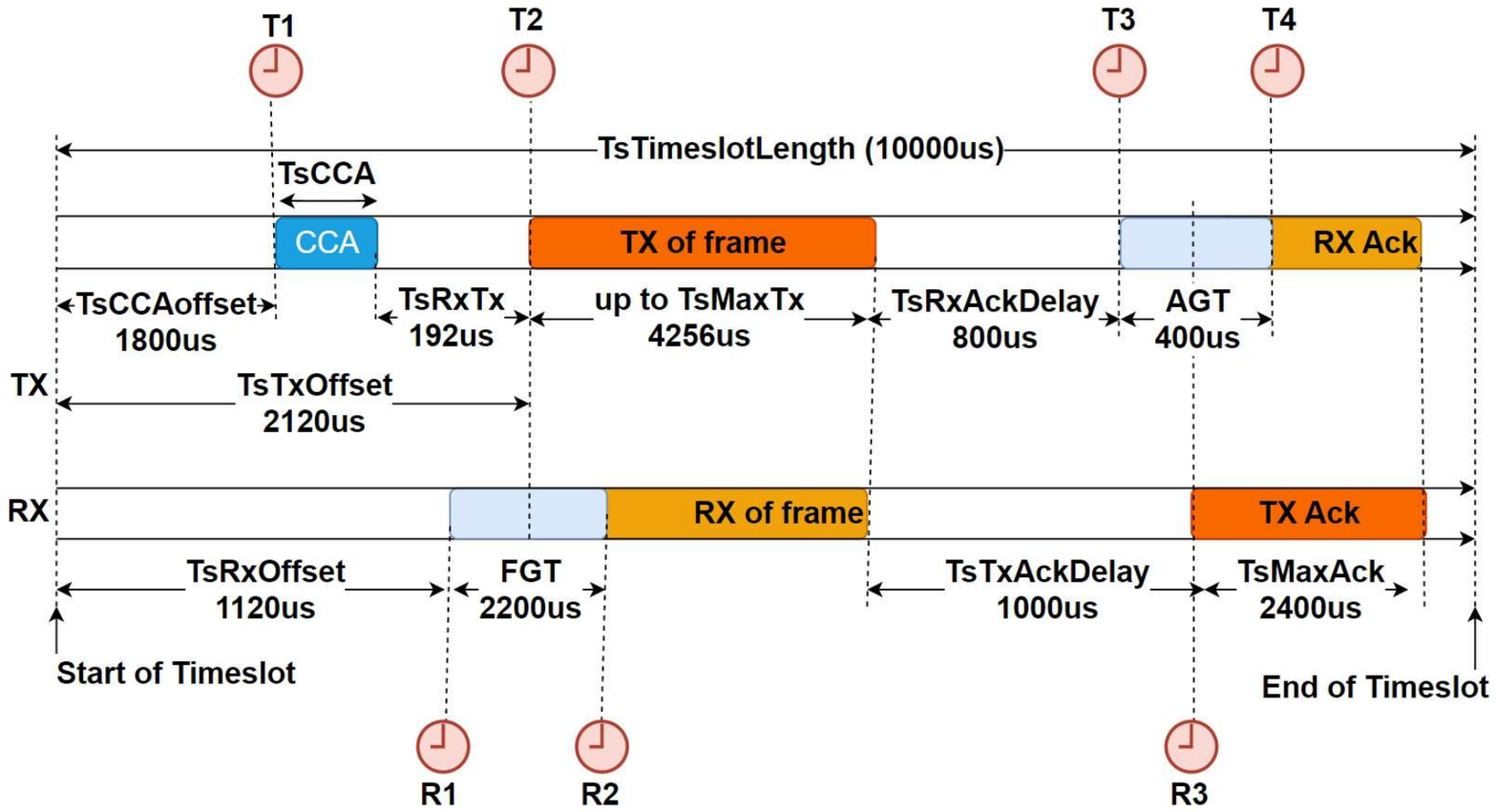}
	\caption{Illustration of slots and channel assignment in TSCH.}
	\label{fig:tsch3}
\end{figure*}

The Time-Slotted Channel Hopping (TSCH) MAC protocol \cite{dujovne20146tisch} is one of the widely used MAC protocols to mitigate the impact of CTI on IEEE 802.15.4 networks. It combines time-division medium access and dynamic channel hopping. In TSCH, a node may transmit a single packet only using a specific channel. Regardless of its success or failure, a transmitter selects a different channel in the next time slot for transmitting the next packet (ref. to Fig. \ref{fig:tsch}). Similarly, a node may transmit a packet only in a time slot assigned to it. It may be assigned multiple time slots in succession. TSCH requires strict time synchronisation to coordinate channel assignment and channel hopping. Fig.~\ref{fig:tsch3} illustrate the importance of time synchronization in TSCH.

At the beginning of a time slot, a transmitter prepares a packet for transmission. The time elapse between the beginning of a time slot and the beginning of a clear channel assessment is called \texttt{TsCCAOffset}. If the channel is free, the transmitter switches to a transmission mode (it is on a receiving mode during clear channel assessment) and sends a packet. Then it switches back to a receiving mode to receive the acknowledgement packet. This packet has to arrive within a fixed amount of time, called Acknowledgement Guard Time or \texttt{AGT}. If it does not arrive, the transmitter considers the packet as a lost packet and awaits the next time slot to re-transmit the packet using a different channel.

\section{The Impact of CTI on Link Quality}
\label{sec:lq}

\begin{figure}[!ht]
	\centering
	\includegraphics[width=0.45\textwidth]{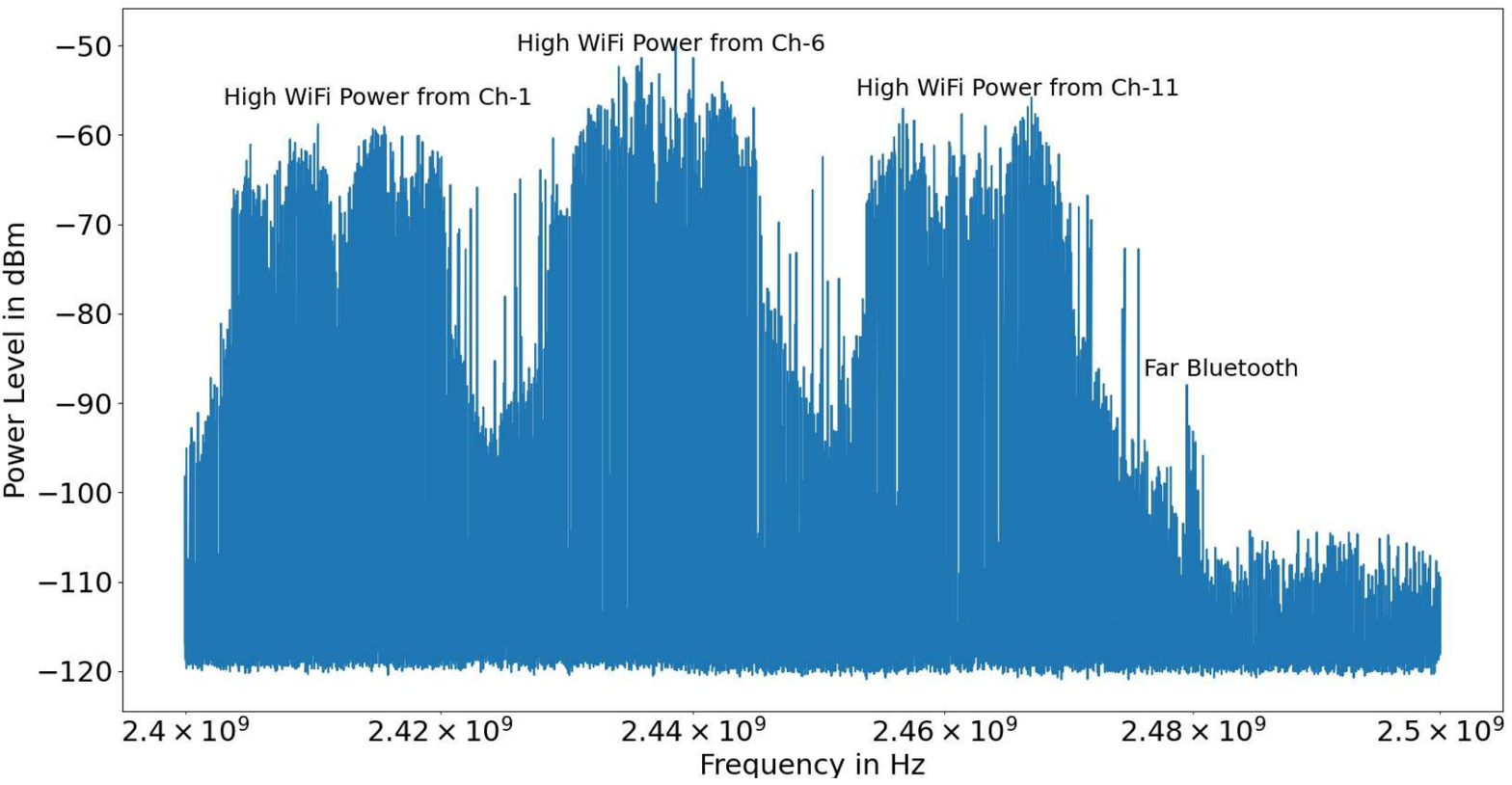}
	\caption{Spectrum occupation by the WiFi network using Channels 1, 6, and 11. The statistics were obtained by scanning the entire ISM Band at a rate of 100 KHz using R\&S PR100 Portable Radio Receiver. Bluetooth and IEEE 802.15.4 utilization are completely masked by the occupation of the three non-overlapping WiFi channels.}
    \label{fig:Blue_wifi_ch1_6_11_spectrum}
    \end{figure}

\begin{figure}[h!]
	\centering
	\includegraphics[width=0.45\textwidth]{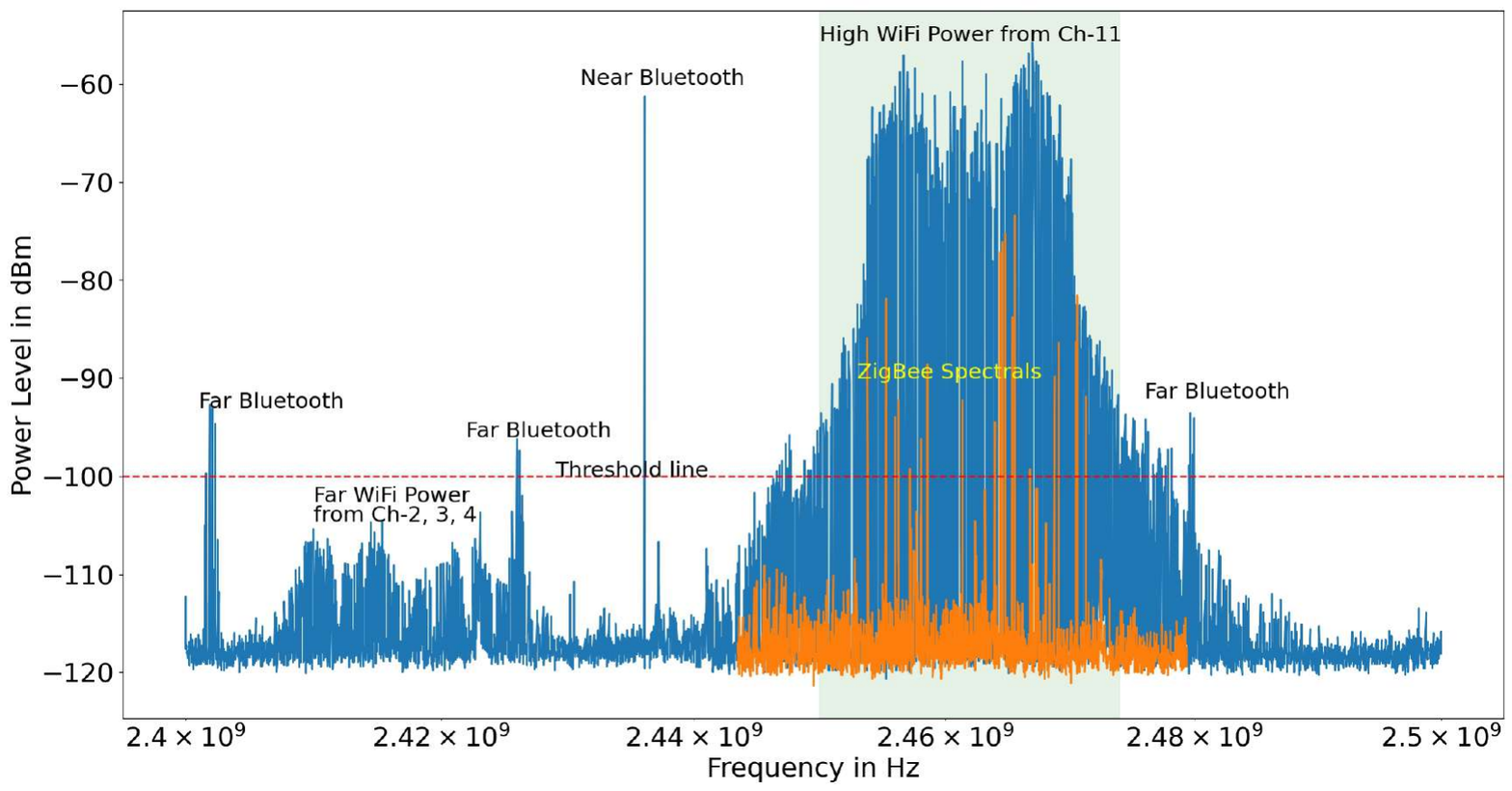}
	\caption{Spectrum occupation by (1) the WiFi network (using Channel 11); (2) an IEEE 802.15.4 (Channels 21, 22, 23, 24) low-power network; and (3) an IEEE 802.15.1 transmitter.}
    \label{fig:Blue_wifi_ch1_11_zigBee_spectrum}
    \end{figure}

Fig. \ref{fig:Blue_wifi_ch1_6_11_spectrum} displays the power spectrum of the three widely used  none-overlapping WiFi channels (1, 6, and 11) when occupied at the same time. The figure clearly shows that the available spectral space between the occupied channels is useful only partially, because the spectral sidelobes extend into this space. It also suggests that the receivers' sensitivity of the low-power network should be appreciably high (above -90 dBm) in order to successfully receive packets destined to them. Fig.  \ref{fig:Blue_wifi_ch1_11_zigBee_spectrum} shows the spectral overlap between Channel 11 of the IEEE 802.11 network and four IEEE 802.15.4 channels (Channels 21-24), when the two networks operate simultaneously. The figure displays the spectrum  of additional, remote networks, which were not a part of the experiment setup. From these two figures it is clear that (1) CTI is a significant challenge for low-power networks and (2) dynamic channel selection should take into consideration the channel utilisation characteristics of both nearby and far away 802.11 networks.

The quality of a wireless link depends on many factors, including the ambient temperature, the noise factor of the receiver, which measures the amount of noise generated inside the receiver, and the signal-to-noise ratio (SNR) of the received signal. The thermal noise power at room temperature (25$^{\circ}$ C or 298 K) in a 1 Hz channel is -204 dBW or -174 dBm/Hz. This value is the reference for any noise power calculation when designing RF systems working at room temperature. The receiver's sensitivity, $\sigma$ (in dBm) takes all these terms into account:
\begin{equation}
   \label{eq:sensitivity}
  \sigma \; (dBm)  = kTB \; (dBm) + SNR \; (dBm) + NF \; (dBm) 
\end{equation}
where $kTB$ is the thermal noise and depends on Boltzmann’s constant, k, in Joules/K, the ambient temperature in degree Kelvin, and the bandwidth of the channel selective filter in the receiver (B). NF is the noise factor of the receiver. For a 2 MHz channel; a SNR of 6 dBm; and a noise factor of 2 dBm, the receiver's sensitivity equals -103 dBm. The sensitivity sets a limit to the minimum SNR of the received signal. In most practical cases, however, a much higher threshold is set to the received power in order to avoid the detection of spurious or corrupted signals.             

A closer examination of the statistics we established from the raw measurements reveal that $93.53\%$ of all the detected signals on the IEEE 802.15.4 Channel 11 (and $94.58\% $, on Channel 12) have RSSI values below the noise floor of the CC2538 radio. In other words, these channels are relatively free of interference. The figure is comparable for Channel 13 and 14; on the former, 4.37\%  and, on the latter, 3.2\% of all the detected signals have a received power exceeding the noise floor of the CC2538 radio. By contrast, the received power detected on Channels  21 to 24 has on average $-60$ dBm, signifying a stark interference. An interesting observation concerns channels 15, 20, 25, and 26. Theoretically, these channels should be the least affected by CTI. In practice, however, this is not the case. The spectral sidelobes of Channels 1, 6, and 11 interfere with the low-power channels. Thus, Channel 15 is affected by the sidelobes of the WiFi Channels 1 and 6; Channel 20 is affected by the sidelobes of the WiFi Channels 6 and 11; and Channels 25 and 26 are affected by the sidelobes of the WiFi Channel 11. From this it can be concluded that the CTI arising from the sidelobes of the IEEE 802.11 channels is considerable. For example, if the detection threshold of the CC2538 radio is $-100$ dBm and the WiFi Channel 11 is active, the interference probability on the low--power channels is as follows: Channel 21 ($97\%$),  Channel 22 ($96.28\%$), Channel 23 ($96.20\%$), and Channel 24 ($96.44\%$).  When the detection threshold is raised to $-83$ dBm,  CTI is reduced significantly, now the interference probability becomes as follows: Channel 21 ($1.94\%$), Channel 22 ($2.51\%$), Channel 23 ($2.67\%$), and Channel 24 ($1.94\%$).

\section{Time Synchronization}
\label{sec:ts}

Time synchronization is a mechanism by which a common time concept amongst the sensor nodes belonging to the same network is established \cite{wu_clock_synch}. For medium access control protocols relying on the assignment of time slots, time synchronization plays a crucial role, since the nodes should have a shared understanding of the beginning and the end of a slot assignment \cite{romanov2020precise}. Similarly, higher layer services and applications often rely on it to function properly \cite{bhushan2020requirements}. 

\begin{figure}[h!]
	\centering
	\includegraphics[width=0.45\textwidth]{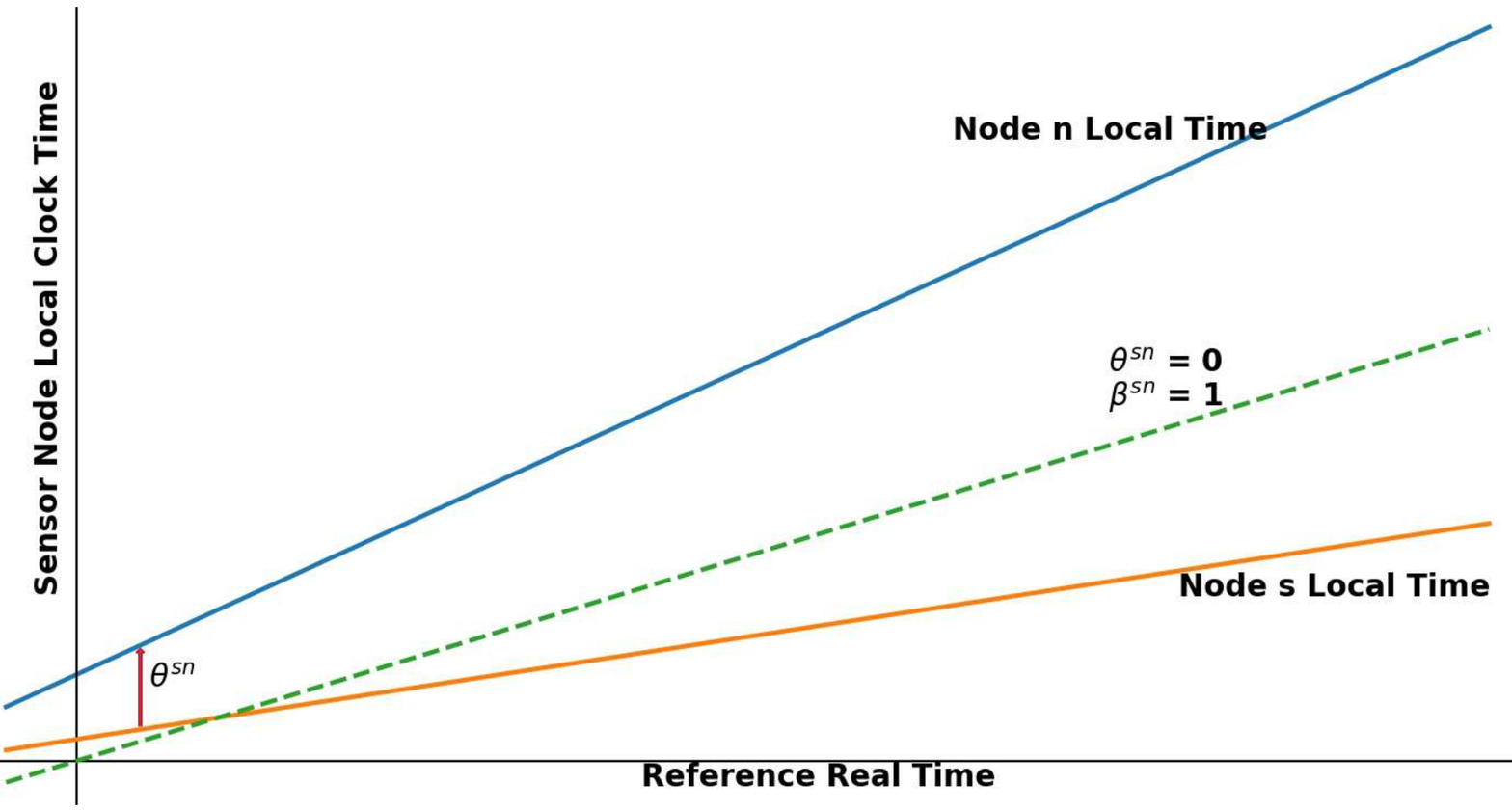}
	\caption{A clock model for two sensing nodes (a sink and a child node) directly communicating.}
\label{fig:clock_model}
\end{figure}

Each individual node in a network maintains its own clock generator, which, due to imperfections in the local crystal oscillator, experiences clock offset and skew. The clock function of a sensor node with respect to a reference time can be expressed as : $C(t) = \theta \ + \beta t$, where $\theta$ is the phase difference or clock offset and $\beta$ is the frequency difference or the clock skew of the crystal oscillator. If one considers the relative clock offset and skew between a child node $n$ and a sink node $s$, the local clock  and its relationship to the clock of the sink can be expressed using equation \ref{equ:clock_offest}:

                \begin{equation}                 
                C_{n}(t) = \theta^{sn}+\beta^{sn}C_{s}(t)                                       
                \label{equ:clock_offest}         
                \end{equation}
where $C_{s}(t)$ is the sink's clock. If the two clocks are perfectly synchronized, the relative phase difference or the relative clock offset is zero ($\theta^{sn} = 0 $ and  $\beta^{sn}=1$), otherwise the child node needs to synchronize its clock using the time information (such as timestamps) it receives from its reference node. Graphically, the relationship between the reference and the local clocks can be illustrated using Fig. \ref{fig:clock_model}. The local clocks increase monotonically in accordance with their clock skew slopes, away from each other when no time synchronization occurs.

When the network is stable and the wireless links are reliable, time synchronization takes place in regular intervals. When this is not the case, clock drifts become nonlinear and the time synchronization intervals are no longer regular. In TSCH, nodes exchange a so-called ``Extended Beacon'' (EB) on a regular basis. This packet contains ``information elements'' -- timing, channel assignments, time slots, slot frames and other preambles and header files. This information is critical for existing nodes to communicate with one another and for new nodes wishing to join the network to learn about channel and slot assignments. 

\begin{figure}[h!]
        \begin{center}      
            \begin{tikzpicture}
                \node at (2,2) (nodeA) {s};
                \node at (2,4) (nodeB) {n};
                \draw[red] (nodeA) circle (2mm);
                \draw[blue] (nodeB) circle (2mm);
                \tikzstyle{arrow} = [thick,->,>=stealth]
                \draw [arrow] (nodeB) -- (9,4);
                \draw [arrow] (nodeA) -- (9,2);
                \draw [arrow] (2.4,4) -- (3.2,2); 
                \draw [arrow] (3.4,2) -- (4.2,4); 
                \draw [arrow] (4.4,4) -- (5.2,2); 
                \draw [arrow] (5.4,2) -- (6.2,4); 
                \draw [decoration={text along path, text={n local time},text align={center}},decorate]  (6.2,4) -- (9,4);
                \draw [decoration={text along path, text={s local time},text align={center}},decorate]  (6.2,2) -- (9,2); 
                \draw [decoration={text along path, text={---},text align={center}},decorate]  (6.2,3) -- (7.2,3); 
                \draw [decoration={text along path, text={T1},text align={center}},decorate]  (2.2,4) -- (2.8,4); 
                \draw [decoration={text along path, text={T2},text align={center}},decorate]  (2.8,1.7) -- (3.4,1.7);
                \draw [decoration={text along path, text={T3},text align={center}},decorate]  (3.4,1.7) -- (3.8,1.7);
                \draw [decoration={text along path, text={T4},text align={center}},decorate]  (3.8,4) -- (4.2,4); 
            \end{tikzpicture}        
        \end{center}
        \caption{ Illustration of a two-way timing message exchange between a sink and a child node during time synchronization measurement: $T_{2}=T_{1}+delay+offset$ and $T_{4}=T_{3}+delay-offset$. The simple time offset between the sink and the child node can be calculated as $[(T_{2}-T_{1})-(T_{4}-T_{3})]/2$}
        \label{fig: two_nodes_time_synch}
    \end{figure}
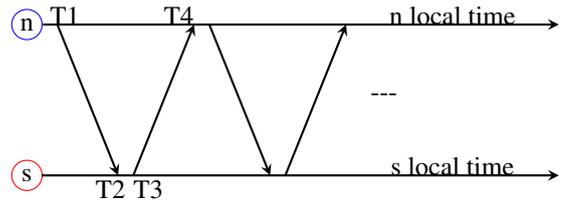

The time deviation between two nodes can be calculated using two different techniques. In the first, the coordinator broadcasts EB packets and a child node determines the  deviation from the timestamps embedded in these packets. In the second (so called ``two-way approach''), the coordinator determines the same from acknowledgement packets and sends the correction along with other timing information to the child  node. This is illustrated in Fig. \ref{fig: two_nodes_time_synch}. Once the network's topology is determined through a higher-level self-organization algorithm, a child node sends its time information to the coordinator at time $T_{1}$; the coordinator node records the arriving time $T_{2}$ and sends a reply  at time $T_{3}$. There is a process delay between  $T_{1}$ and  $T_{3}$. This delay is included in $T_{3}$.  Finally, the child node records the time at which it receive the correction packet at time stamp $T_{4}$. 

Using Equation \ref{equ:clock_offest} and Fig.\ref{fig: two_nodes_time_synch}, the local time of the coordinator and the child node can be determined as follows \cite{wu_clock_synch},\cite{Shi_interference_wifizigBee_2017}:

   \begin{equation}
                 \begin{split}
                    T_{2} = \theta+\beta(T_{1} + \tau + \delta ): \ client \ time \ stamp \\
                    T_{3} = \theta+\beta(T_{4} - \tau - \sigma ): \ sink \ time \ stamp  \\                   
                 \end{split}                     
                \label{equ:timestamp_clock_offest}         
                \end{equation}
Where $\beta$ and  $\theta$ the are relative clock skew and offset, respectively, of the child node with respect to the coordinator; $\tau $ is a fixed network delay; $\delta$ and  $\sigma$ are variables related to transmission delay (send, receive, process, etc.) at the coordinator and the child node, respectively. The three parameters -- $\beta$, $\theta$ and $\tau $ -- are factors, which determine time synchronization frequency. When there is a significant packet loss, their values increase and the nodes make frequent but unsuccessful attempts to synchronise time.               

      \begin{figure}[h!]
            	\centering
            	\includegraphics[width=0.45\textwidth]{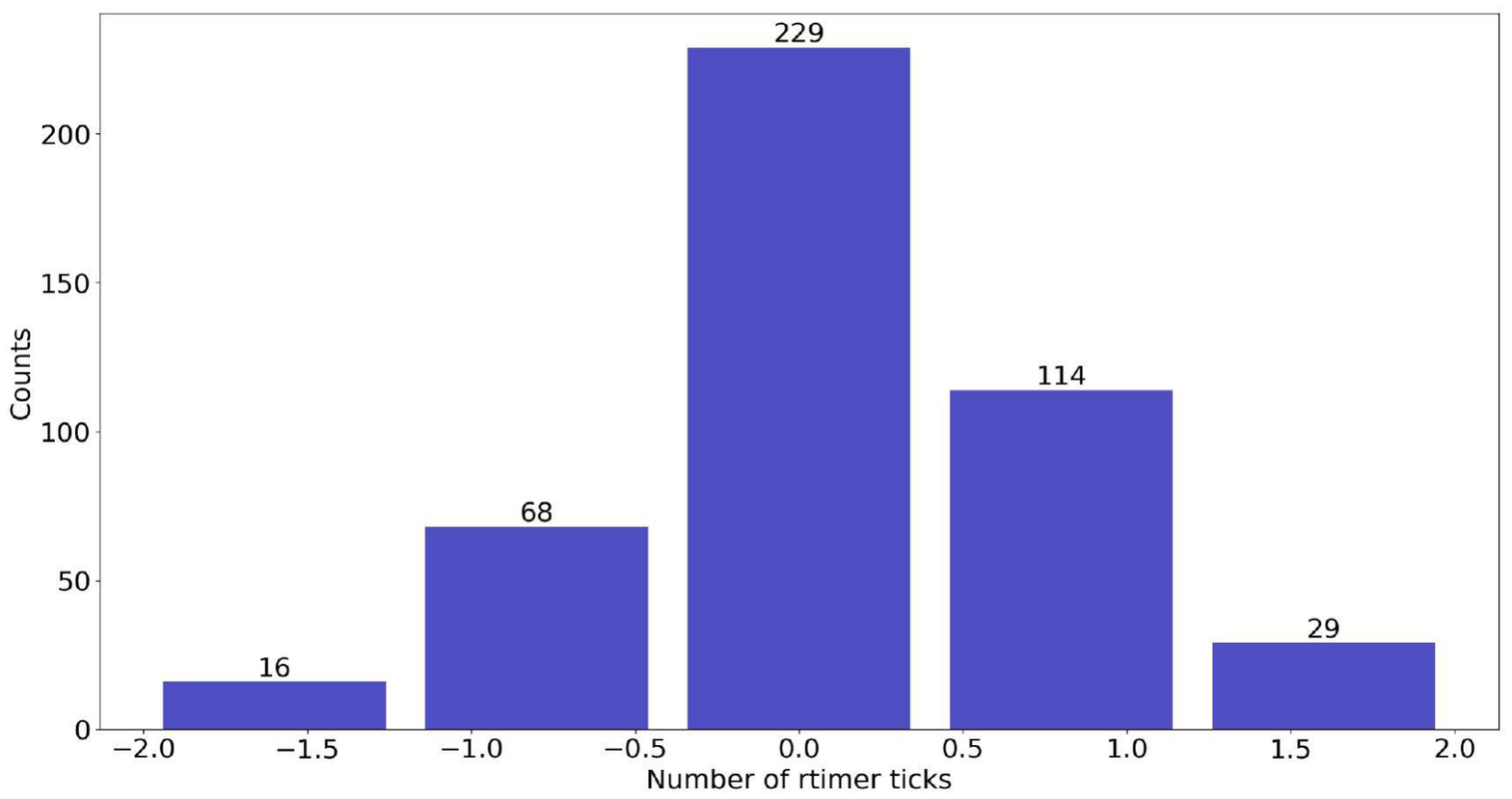}
            	\caption{The histogram of the time drift of a child node with respect to a sink node between two time synchronisation intervals. The statistics are established for an observation duration of 1 hour. The WiFi network producing the CTI was occupying Channel 11. The interference reaching the low-power network had an average magnitude of $-83$ dBm.} 
            \label{fig: time_drift_interference_1}
            \end{figure}

\begin{figure}[h!]
    \centering
    \includegraphics[width=0.45\textwidth]{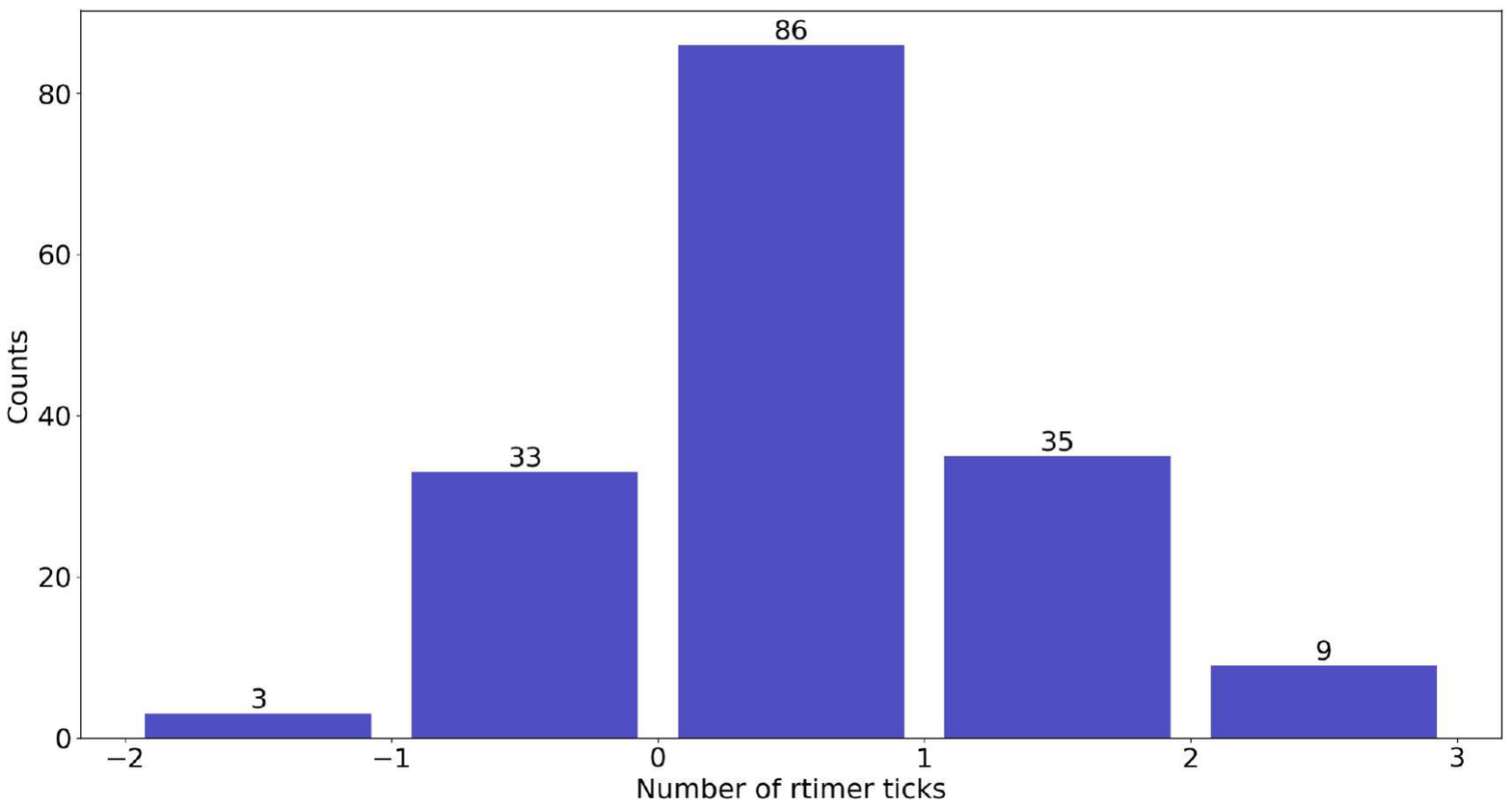} 
            	\caption{The histogram of the time drift of a child node with respect to a sink node between two time synchronisation intervals. The statistics are established for an observation duration of 1 hour. The WiFi network producing the CTI was occupying Channel 11. The interference reaching the low-power network had an average magnitude of $-30$ dBm.} 
    \label{fig: time_drift_interference_2}
\end{figure}

The Contiki operating system offers a set of timer libraries. The underlying clock module provides system time as a 32-bit unsigned integer. Time drift corrections are made in terms of clock ticks at a drift rate at ppm and are expressed as a 32 bit integer. Typically 1 clock tick can vary between 1 and 10 milliseconds. TSCH sends out Extended Beacon (EB) every second. In the absence of a considerable CTI, the relative time drift between two synchronization attempts is approximately 1 rtimer tick (between $-1$ and $1$ ticks). In order to investigate the impact of CTI, we considered two power levels, namely, a high CTI, $-30$ dBm, and a low CTI,  $-86$ dBm. In the case of a high CTI, the time drift between two successful synchronization attempts increases, now extending between $-3$ and $3$ ticks. Fig.~ \ref{fig: time_drift_interference_1} shows the time drift histogram between the child nodes and the coordinator during a 1 hour observation window. In the case of a low CTI, the time drift distribution exhibited a slight improvement. Even though the shape of the distribution is similar to the previous case, the variance becomes smaller, as can be seen in Fig.~\ref{fig: time_drift_interference_2}. Hence, in the case of a low CTI, $P\{\mathbf{c} \geq 61.02 \ us\} = 0.07$, whereas in the case of a high CTI, $P\{\mathbf{c} \geq 61.02 \ us\} = 0.1$, where $\mathbf{c}$ is the relative clock drift regarded as a random variable. Similarly, in the case of a modest CTI, $P\{\mathbf{c} \leq 30.5 \ us\} = 0.41$, whereas in the case of a high CTI,  $P\{\mathbf{c} \leq 30.5 \ us\} = 0.39$. In general, in the presence of CTI, on average, $36.40\%$ additional synchronization messages were exchanged in 1 hour in an attempt to keep the relative time drift within $\pm 1$ rtimer tick. The more synchronization massages are exchanged, the higher is the energy cost of time synchronization. 

\section{Network Join Time}
\label{sec:latency}
One of the desirable qualities of wireless sensor networks is self-organisation: Nodes can independently establish a multi-hop network; new nodes can join the network; and nodes, which leave the network for various reasons, can rejoin. The MAC protocol plays a key role during self-organisation. As stated above, in TSCH, the coordinator node broadcasts EB packets regularly. A node wishing to join the network listens to these packets, and upon receiving one, contends for the medium and sends a request-to-join packet using a unicast channel. Because initially no time synchronisation does take place between the coordinator and the nodes joining the network, almost certainly there is a time difference between the coordinator and these nodes. Expecting this condition, TSCH defines two types of time offsets. The first offset is intended to prevent a receiver from early sleeping in case a packet (preamble, SFD, Headers, payload, FCS) does not arrive according to the receiver's local time. The second time offset is intended for a transmitter to receive a delayed acknowledgement packet.

    \begin{figure}[h!]
        \centering
        \includegraphics[width=0.45\textwidth]{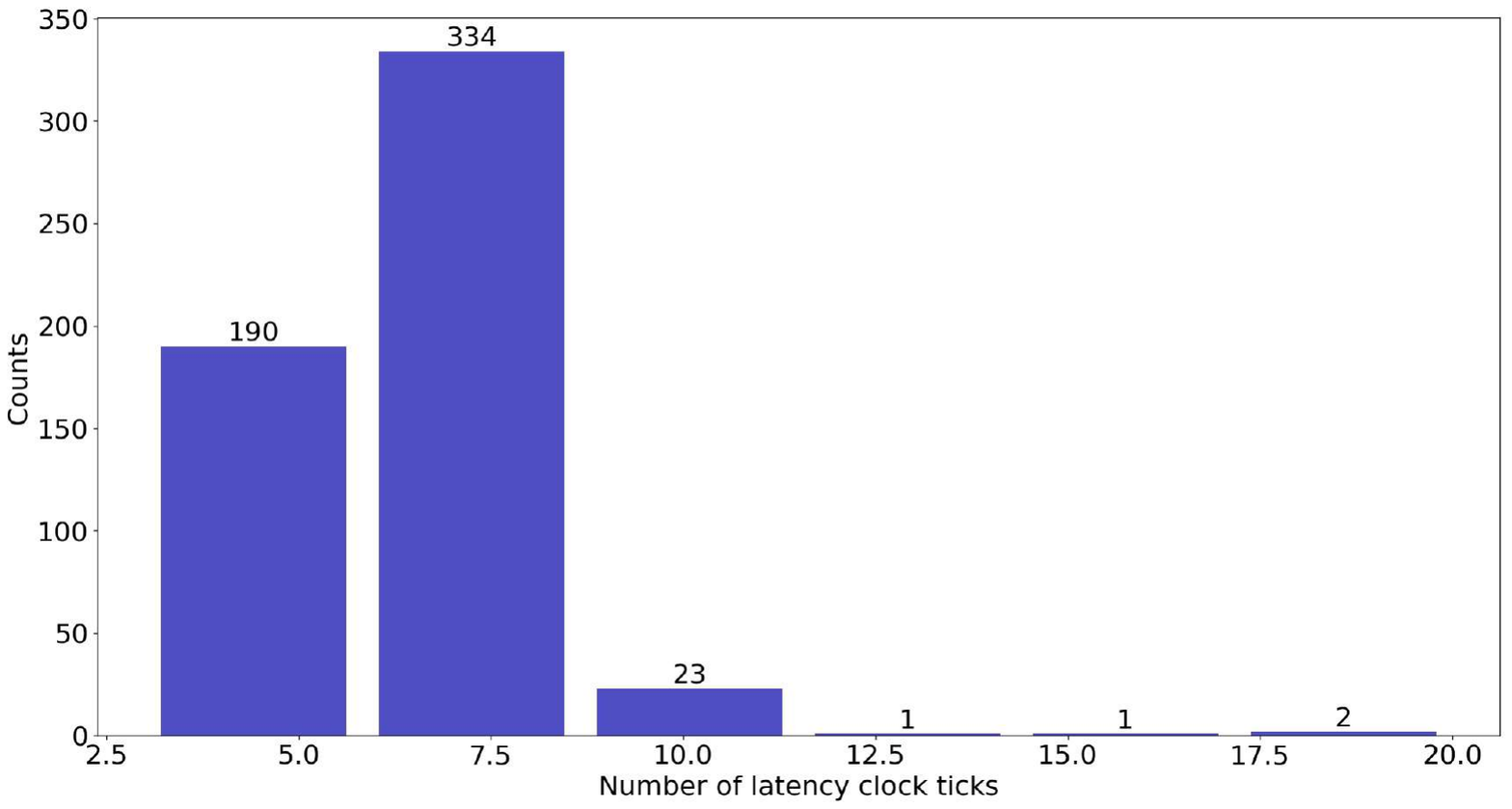}
        \caption{The histogram of network join latency during a moderate CTI.}
        \label{fig: latency_with_interference}
    \end{figure}

    \begin{figure}[h!]
        \centering
        \includegraphics[width=0.45\textwidth]{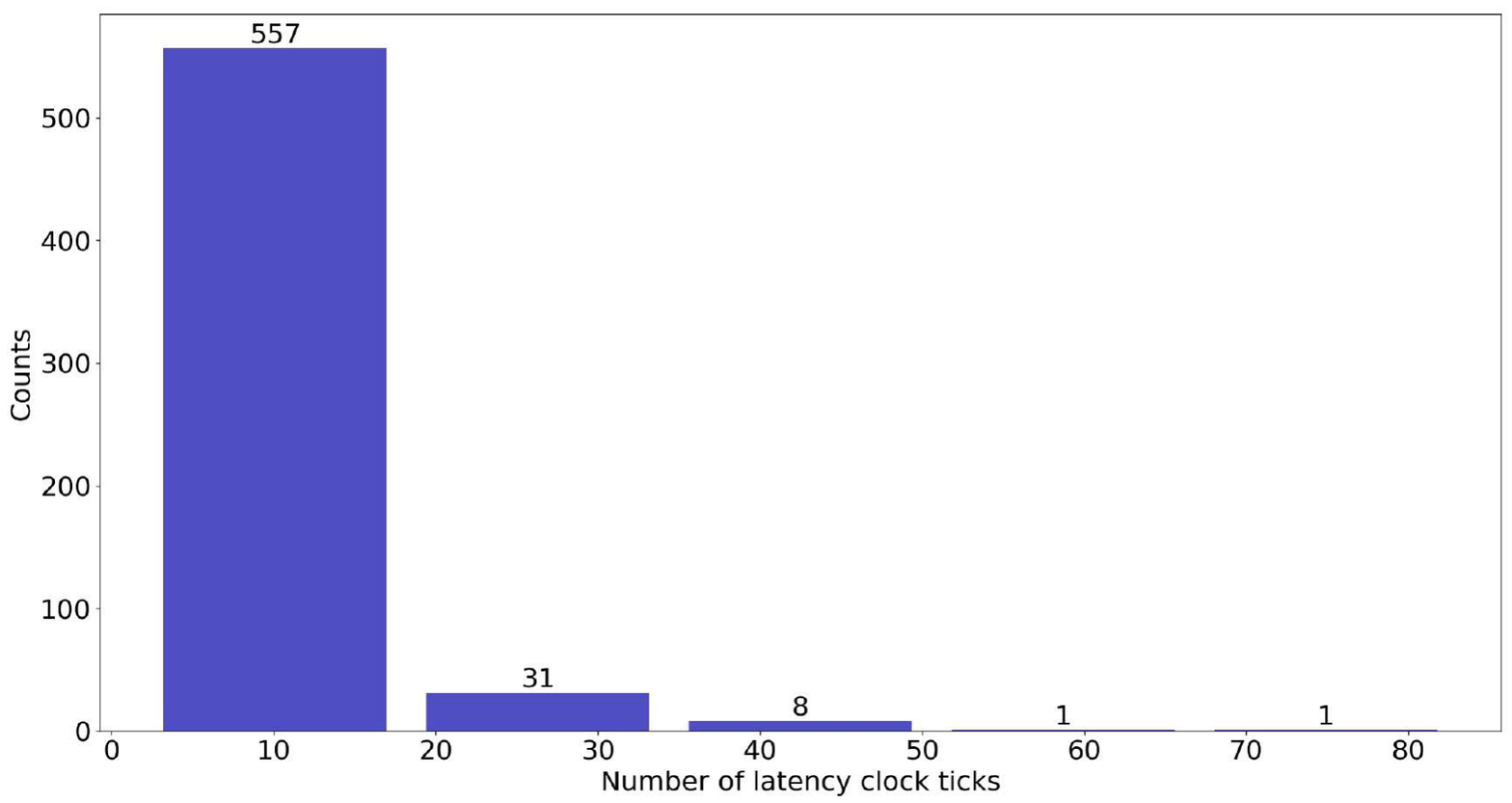}
        \caption{The histogram of network join latency during an high CTI.}
        \label{fig: latency_no_interference}
    \end{figure}  

When the network is under the influence of a CTI, as we already discussed in Section~\ref{sec:ts}, the time drifts between the coordinator and the new nodes increase and the two time offsets are not sufficient to establish a reliable communication. This creates a join delay. We measured the join delay with and without a CTI. The distributions of these delay are given in Figs. \ref{fig: latency_with_interference} and \ref{fig: latency_no_interference}.   Without CTI, $83.3\%$  of the case, the join delay  is between $40$ and $70$ ms, whereas in the presence of CTI, $96.82\%$ of the time, the join delay is between $100$ and $200$ ms. In other words, the join delay in the presence of CTI is about five times higher than without CTI. Additionally, the maximum join delay we observed during CTI was $800$ ms, whereas it was $200$ ms when there was no CTI.     
\section{Conclusion}
\label{sec:conclusion}

In this paper we investigated the impact of CTI on time synchronization and network join time of low-power sensing nodes. Medium Access Control protocols which are designed to deal with CTI often rely on time synchronization and dynamic channel hopping. In the presence of CTI, however, packets containing synchronization as well as slot and channel assignment information may be subject to corruption. This condition further exacerbates slot and channel occupation and causes further mutual interference. The study shows that in the absence of a CTI, the clock drift between two synchronization intervals (1 s) is within $\pm 1$ clock tick. when the average CTI reaching a low-power network is about $-86$ dBm, i.e., a moderate CTI, the clock drift between two synchronization intervals is between $\pm 2$ clock ticks. When the CTI magnitude increases to $-30$ dBm, the clock drift increases to $\pm 3$ clock ticks. The impact of CTI is even worse on network join latency. Nodes have to receive a join invitation from a coordinator. When the network relies on the Time-Slotted Channel-Hopping (TSCH) protocol, the invitation comes in the form of an Extended Beacon, which, a coordinator node broadcasts every second. For our case, all nodes receive this invitation directly. In the absence of a CTI, the join latency is below 5 clock ticks. In the presence of a moderate CTI, the latency increases to about 7  clock ticks. When the CTI becomes considerable, network join time increases considerably as well, now exceeding 10 clock ticks on average. Our future work is to further address this issue and improve time synchronization in low-power networks operating in the presence of CTI. 
\section{Acknowledgment}
The field experiments presented in this paper were carried out in collaboration with colleagues at Knight Foundation School of Computing and Information Sciences and Institute of Environment, Florida International University. This includes Professor Christian Poellabauer and his team; Professor Leonardo Bobadilla and his team; and Prorfessor Camilo Roa Penarete and his team.

\balance
\bibliographystyle{elsarticle-harv} 
\bibliography{library}

\begin{thebibliography}{55}
\expandafter\ifx\csname natexlab\endcsname\relax\def\natexlab#1{#1}\fi
\providecommand{\url}[1]{\texttt{#1}}
\providecommand{\href}[2]{#2}
\providecommand{\path}[1]{#1}
\providecommand{\DOIprefix}{doi:}
\providecommand{\ArXivprefix}{arXiv:}
\providecommand{\URLprefix}{URL: }
\providecommand{\Pubmedprefix}{pmid:}
\providecommand{\doi}[1]{\href{http://dx.doi.org/#1}{\path{#1}}}
\providecommand{\Pubmed}[1]{\href{pmid:#1}{\path{#1}}}
\providecommand{\bibinfo}[2]{#2}
\ifx\xfnm\relax \def\xfnm[#1]{\unskip,\space#1}\fi
\bibitem[{Athanasios~Papoulis(2001)}]{Papoulis_probability_2001}
\bibinfo{author}{Athanasios~Papoulis, S.U.P.}, \bibinfo{year}{2001}.
\newblock \bibinfo{title}{Probability, Random Variables and Stochastic Processes, 4th edition}.
\newblock \bibinfo{publisher}{McGraw-Hill Higher Education}.
\bibitem[{Bhushan and Sahoo(2020)}]{bhushan2020requirements}
\bibinfo{author}{Bhushan, B.}, \bibinfo{author}{Sahoo, G.}, \bibinfo{year}{2020}.
\newblock \bibinfo{title}{Requirements, protocols, and security challenges in wireless sensor networks: An industrial perspective}.
\newblock \bibinfo{journal}{Handbook of computer networks and cyber security: principles and paradigms} , \bibinfo{pages}{683--713}.
\bibitem[{Booranawong et~al.(2021)Booranawong, Sengchuai, Buranapanichkit, Jindapetch and Saito}]{Booranawong_Rssi_indoorlocalization_2021}
\bibinfo{author}{Booranawong, A.}, \bibinfo{author}{Sengchuai, K.}, \bibinfo{author}{Buranapanichkit, D.}, \bibinfo{author}{Jindapetch, N.}, \bibinfo{author}{Saito, H.}, \bibinfo{year}{2021}.
\newblock \bibinfo{title}{Rssi-based indoor localization using multi-lateration with zone selection and virtual position-based compensation methods}.
\newblock \bibinfo{journal}{IEEE Access} \bibinfo{volume}{9}, \bibinfo{pages}{46223--46239}.
\newblock \DOIprefix\doi{10.1109/ACCESS.2021.3068295}.
\bibitem[{Brown et~al.(2014)Brown, Roedig, Boano and Römer}]{brown_PRR_2014}
\bibinfo{author}{Brown, J.}, \bibinfo{author}{Roedig, U.}, \bibinfo{author}{Boano, C.A.}, \bibinfo{author}{Römer, K.}, \bibinfo{year}{2014}.
\newblock \bibinfo{title}{Estimating packet reception rate in noisy environments}, in: \bibinfo{booktitle}{39th Annual IEEE Conference on Local Computer Networks Workshops}, pp. \bibinfo{pages}{583--591}.
\newblock \DOIprefix\doi{10.1109/LCNW.2014.6927706}.
\bibitem[{CC2538™(2013)}]{cc2538_soc_2013}
\bibinfo{author}{CC2538™, T.I.}, \bibinfo{year}{2013}.
\newblock \bibinfo{title}{Cc2538 system-on-chip solution for 2.4-ghz ieee 802.15.4 and zigbee®/zigbee ip® applications}.
\newblock \URLprefix \url{https://www.ti.com/lit/ug/swru319c/swru319c.pdf?ts=1706703279258}.
\bibitem[{Chang et~al.(2015)Chang, Watteyne, Pister and Wang}]{chang2015adaptive}
\bibinfo{author}{Chang, T.}, \bibinfo{author}{Watteyne, T.}, \bibinfo{author}{Pister, K.}, \bibinfo{author}{Wang, Q.}, \bibinfo{year}{2015}.
\newblock \bibinfo{title}{Adaptive synchronization in multi-hop tsch networks}.
\newblock \bibinfo{journal}{Computer Networks} \bibinfo{volume}{76}, \bibinfo{pages}{165--176}.
\bibitem[{Chi et~al.(2019)Chi, Li, Sun, Yao and Zhu}]{chi2019concurrent}
\bibinfo{author}{Chi, Z.}, \bibinfo{author}{Li, Y.}, \bibinfo{author}{Sun, H.}, \bibinfo{author}{Yao, Y.}, \bibinfo{author}{Zhu, T.}, \bibinfo{year}{2019}.
\newblock \bibinfo{title}{Concurrent cross-technology communication among heterogeneous iot devices}.
\newblock \bibinfo{journal}{IEEE/ACM Transactions on Networking} \bibinfo{volume}{27}, \bibinfo{pages}{932--947}.
\bibitem[{Contiki-NG(2018-2022a)}]{contiki_NG_CoAP}
\bibinfo{author}{Contiki-NG}, \bibinfo{year}{2018-2022}a.
\newblock \bibinfo{title}{Coap in contiki-ng}.
\newblock \URLprefix \url{https://docs.contiki-ng.org/en/develop/doc/programming/CoAP.html}.
\bibitem[{Contiki-NG(2018-2022b)}]{contiki_NG_Rtimer}
\bibinfo{author}{Contiki-NG}, \bibinfo{year}{2018-2022}b.
\newblock \bibinfo{title}{Timers in contiki-ng}.
\newblock \URLprefix \url{https://docs.contiki-ng.org/en/develop/doc/programming/Timers.html#the-rtimer-libraryl}.
\bibitem[{Dargie and Kidane(2024)}]{dargie2024mitigating}
\bibinfo{author}{Dargie, W.}, \bibinfo{author}{Kidane, Z.M.}, \bibinfo{year}{2024}.
\newblock \bibinfo{title}{Mitigating cross-technology interference in low-power wireless networks}, in: \bibinfo{booktitle}{2024 33rd International Conference on Computer Communications and Networks (ICCCN)}, \bibinfo{organization}{IEEE}. pp. \bibinfo{pages}{1--8}.
\bibitem[{Dolha et~al.(2019)Dolha, Negirla, Alexa and Silea}]{Dolha_signallevel_2019}
\bibinfo{author}{Dolha, S.}, \bibinfo{author}{Negirla, P.}, \bibinfo{author}{Alexa, F.}, \bibinfo{author}{Silea, I.}, \bibinfo{year}{2019}.
\newblock \bibinfo{title}{Considerations about the signal level measurement in wireless sensor networks for node position estimation}.
\newblock \bibinfo{journal}{Sensors} \bibinfo{volume}{19}.
\newblock \URLprefix \url{https://www.mdpi.com/1424-8220/19/19/4179}, \DOIprefix\doi{10.3390/s19194179}.
\bibitem[{Dujovne et~al.(2014)Dujovne, Watteyne, Vilajosana and Thubert}]{dujovne20146tisch}
\bibinfo{author}{Dujovne, D.}, \bibinfo{author}{Watteyne, T.}, \bibinfo{author}{Vilajosana, X.}, \bibinfo{author}{Thubert, P.}, \bibinfo{year}{2014}.
\newblock \bibinfo{title}{6tisch: deterministic ip-enabled industrial internet (of things)}.
\newblock \bibinfo{journal}{IEEE Communications Magazine} \bibinfo{volume}{52}, \bibinfo{pages}{36--41}.
\bibitem[{Duquennoy et~al.(2017)Duquennoy, Elsts, Al~Nahas and Oikonomo}]{duquennoy2017tsch}
\bibinfo{author}{Duquennoy, S.}, \bibinfo{author}{Elsts, A.}, \bibinfo{author}{Al~Nahas, B.}, \bibinfo{author}{Oikonomo, G.}, \bibinfo{year}{2017}.
\newblock \bibinfo{title}{Tsch and 6tisch for contiki: Challenges, design and evaluation}, in: \bibinfo{booktitle}{2017 13th International Conference on Distributed Computing in Sensor Systems (DCOSS)}, \bibinfo{organization}{IEEE}. pp. \bibinfo{pages}{11--18}.
\bibitem[{Elias et~al.(2014)Elias, Paris and Krunz}]{elias2014cross}
\bibinfo{author}{Elias, J.}, \bibinfo{author}{Paris, S.}, \bibinfo{author}{Krunz, M.}, \bibinfo{year}{2014}.
\newblock \bibinfo{title}{Cross-technology interference mitigation in body area networks: An optimization approach}.
\newblock \bibinfo{journal}{IEEE Transactions on Vehicular Technology} \bibinfo{volume}{64}, \bibinfo{pages}{4144--4157}.
\bibitem[{Gao et~al.(2023)Gao, Liu, Hu, Wang, Chen, Chen and He}]{Gao_tsynch_cti_2023}
\bibinfo{author}{Gao, D.}, \bibinfo{author}{Liu, Y.}, \bibinfo{author}{Hu, B.}, \bibinfo{author}{Wang, L.}, \bibinfo{author}{Chen, W.}, \bibinfo{author}{Chen, Y.}, \bibinfo{author}{He, T.}, \bibinfo{year}{2023}.
\newblock \bibinfo{title}{Time synchronization based on cross-technology communication for iot networks}.
\newblock \bibinfo{journal}{IEEE Internet of Things Journal} \bibinfo{volume}{10}, \bibinfo{pages}{19753--19764}.
\newblock \DOIprefix\doi{10.1109/JIOT.2023.3282202}.
\bibitem[{Gaotao~Shi(2017)}]{Shi_interference_wifizigBee_2017}
\bibinfo{author}{Gaotao~Shi, K.L.}, \bibinfo{year}{2017}.
\newblock \bibinfo{title}{Signal Interference in WiFi and ZigBee Networks}.
\newblock \bibinfo{publisher}{Spinger}.
\bibitem[{Golmie et~al.(2003)Golmie, R{\'e}bala and Chevrollier}]{golmie2003bluetooth}
\bibinfo{author}{Golmie, N.}, \bibinfo{author}{R{\'e}bala, O.}, \bibinfo{author}{Chevrollier, N.}, \bibinfo{year}{2003}.
\newblock \bibinfo{title}{Bluetooth adaptive frequency hopping and scheduling}, in: \bibinfo{booktitle}{IEEE Military Communications Conference, 2003. MILCOM 2003.}, \bibinfo{organization}{IEEE}. pp. \bibinfo{pages}{1138--1142}.
\bibitem[{Grimaldi et~al.(2020)Grimaldi, Mahmood, Hassan, Gidlund and Hancke}]{grimaldi2020autonomous}
\bibinfo{author}{Grimaldi, S.}, \bibinfo{author}{Mahmood, A.}, \bibinfo{author}{Hassan, S.A.}, \bibinfo{author}{Gidlund, M.}, \bibinfo{author}{Hancke, G.P.}, \bibinfo{year}{2020}.
\newblock \bibinfo{title}{Autonomous interference mapping for industrial internet of things networks over unlicensed bands: Identifying cross-technology interference}.
\newblock \bibinfo{journal}{IEEE Industrial Electronics Magazine} \bibinfo{volume}{15}, \bibinfo{pages}{67--78}.
\bibitem[{Guo et~al.(2020a)Guo, Li, Zhao, Wen and Tang}]{guo_imperfect_synchronization_2020}
\bibinfo{author}{Guo, W.}, \bibinfo{author}{Li, C.}, \bibinfo{author}{Zhao, H.}, \bibinfo{author}{Wen, R.}, \bibinfo{author}{Tang, Y.}, \bibinfo{year}{2020}a.
\newblock \bibinfo{title}{Comprehensive effects of imperfect synchronization and channel estimation on known interference cancellation}.
\newblock \bibinfo{journal}{IEEE Transactions on Vehicular Technology} \bibinfo{volume}{69}, \bibinfo{pages}{457--470}.
\newblock \DOIprefix\doi{10.1109/TVT.2019.2950046}.
\bibitem[{Guo et~al.(2020b)Guo, He, Zheng, Yu and Gnawali}]{guo2020zigfi}
\bibinfo{author}{Guo, X.}, \bibinfo{author}{He, Y.}, \bibinfo{author}{Zheng, X.}, \bibinfo{author}{Yu, L.}, \bibinfo{author}{Gnawali, O.}, \bibinfo{year}{2020}b.
\newblock \bibinfo{title}{Zigfi: Harnessing channel state information for cross-technology communication}.
\newblock \bibinfo{journal}{IEEE/ACM Transactions on Networking} \bibinfo{volume}{28}, \bibinfo{pages}{301--311}.
\bibitem[{Hermeto et~al.(2017)Hermeto, Gallais and Theoleyre}]{hermeto2017scheduling}
\bibinfo{author}{Hermeto, R.T.}, \bibinfo{author}{Gallais, A.}, \bibinfo{author}{Theoleyre, F.}, \bibinfo{year}{2017}.
\newblock \bibinfo{title}{Scheduling for ieee802. 15.4-tsch and slow channel hopping mac in low power industrial wireless networks: A survey}.
\newblock \bibinfo{journal}{Computer Communications} \bibinfo{volume}{114}, \bibinfo{pages}{84--105}.
\bibitem[{Hithnawi et~al.(2016)Hithnawi, Li, Shafagh, Gross and Duquennoy}]{hithnawi2016crosszig}
\bibinfo{author}{Hithnawi, A.}, \bibinfo{author}{Li, S.}, \bibinfo{author}{Shafagh, H.}, \bibinfo{author}{Gross, J.}, \bibinfo{author}{Duquennoy, S.}, \bibinfo{year}{2016}.
\newblock \bibinfo{title}{Crosszig: Combating cross-technology interference in low-power wireless networks}, in: \bibinfo{booktitle}{2016 15th ACM/IEEE International Conference on Information Processing in Sensor Networks (IPSN)}, \bibinfo{organization}{Ieee}. pp. \bibinfo{pages}{1--12}.
\bibitem[{Iova et~al.(2016)Iova, Picco, Istomin and Kiraly}]{iova2016rpl}
\bibinfo{author}{Iova, O.}, \bibinfo{author}{Picco, P.}, \bibinfo{author}{Istomin, T.}, \bibinfo{author}{Kiraly, C.}, \bibinfo{year}{2016}.
\newblock \bibinfo{title}{Rpl: The routing standard for the internet of things... or is it?}
\newblock \bibinfo{journal}{IEEE Communications Magazine} \bibinfo{volume}{54}, \bibinfo{pages}{16--22}.
\bibitem[{Jais et~al.(2016)Jais, Sabapathy, Jusoh, Ehkan, Murukesan, Ismail and Ahmad}]{Jais_rssi_code_2016}
\bibinfo{author}{Jais, M.I.}, \bibinfo{author}{Sabapathy, T.}, \bibinfo{author}{Jusoh, M.}, \bibinfo{author}{Ehkan, P.}, \bibinfo{author}{Murukesan, L.}, \bibinfo{author}{Ismail, I.}, \bibinfo{author}{Ahmad, R.B.}, \bibinfo{year}{2016}.
\newblock \bibinfo{title}{Received signal strength indication (rssi) code assessment for wireless sensors network (wsn) deployed raspberry-pi}, in: \bibinfo{booktitle}{2016 International Conference on Robotics, Automation and Sciences (ICORAS)}, pp. \bibinfo{pages}{1--4}.
\newblock \DOIprefix\doi{10.1109/ICORAS.2016.7872618}.
\bibitem[{Kim and He(2015)}]{kim2015freebee}
\bibinfo{author}{Kim, S.M.}, \bibinfo{author}{He, T.}, \bibinfo{year}{2015}.
\newblock \bibinfo{title}{Freebee: Cross-technology communication via free side-channel}, in: \bibinfo{booktitle}{Proceedings of the 21st Annual International Conference on Mobile Computing and Networking}, pp. \bibinfo{pages}{317--330}.
\bibitem[{Konings et~al.(2017)Konings, Faulkner, Alam, Noble and Lai}]{Konings_effects_interferenec_2017}
\bibinfo{author}{Konings, D.}, \bibinfo{author}{Faulkner, N.}, \bibinfo{author}{Alam, F.}, \bibinfo{author}{Noble, F.}, \bibinfo{author}{Lai, E.M.K.}, \bibinfo{year}{2017}.
\newblock \bibinfo{title}{The effects of interference on the rssi values of a zigbee based indoor localization system}, in: \bibinfo{booktitle}{2017 24th International Conference on Mechatronics and Machine Vision in Practice (M2VIP)}, pp. \bibinfo{pages}{1--5}.
\newblock \DOIprefix\doi{10.1109/M2VIP.2017.8211460}.
\bibitem[{Li et~al.(2022)Li, Kim, Zhang, Huan and Smith}]{Li-Sihao_delaytsch_2022}
\bibinfo{author}{Li, S.}, \bibinfo{author}{Kim, K.S.}, \bibinfo{author}{Zhang, L.}, \bibinfo{author}{Huan, X.}, \bibinfo{author}{Smith, J.}, \bibinfo{year}{2022}.
\newblock \bibinfo{title}{Energy-efficient message bundling with delay and synchronization constraints in wireless sensor networks}.
\newblock \bibinfo{journal}{Sensors} \bibinfo{volume}{22}.
\newblock \URLprefix \url{https://www.mdpi.com/1424-8220/22/14/5276}, \DOIprefix\doi{10.3390/s22145276}.
\bibitem[{Li and He(2017)}]{li2017webee}
\bibinfo{author}{Li, Z.}, \bibinfo{author}{He, T.}, \bibinfo{year}{2017}.
\newblock \bibinfo{title}{Webee: Physical-layer cross-technology communication via emulation}, in: \bibinfo{booktitle}{Proceedings of the 23rd Annual International Conference on Mobile Computing and Networking}, pp. \bibinfo{pages}{2--14}.
\bibitem[{Mao et~al.(2023)Mao, Zhao, Xia, Yang, Xu, Liu and Huang}]{mao_link_quality_2023}
\bibinfo{author}{Mao, J.}, \bibinfo{author}{Zhao, Y.}, \bibinfo{author}{Xia, Y.}, \bibinfo{author}{Yang, Z.}, \bibinfo{author}{Xu, C.}, \bibinfo{author}{Liu, W.}, \bibinfo{author}{Huang, D.}, \bibinfo{year}{2023}.
\newblock \bibinfo{title}{Revisiting link quality metrics and models for multichannel low-power lossy networks}.
\newblock \bibinfo{journal}{Sensors} \bibinfo{volume}{23}.
\newblock \URLprefix \url{https://www.mdpi.com/1424-8220/23/3/1303}, \DOIprefix\doi{10.3390/s23031303}.
\bibitem[{Marquez et~al.(2020)Marquez, Osorio, Calle, Velez, Serrano and Candelo-Becerra}]{Luz_lorawan_block_interference_2020}
\bibinfo{author}{Marquez, L.E.}, \bibinfo{author}{Osorio, A.}, \bibinfo{author}{Calle, M.}, \bibinfo{author}{Velez, J.C.}, \bibinfo{author}{Serrano, A.}, \bibinfo{author}{Candelo-Becerra, J.E.}, \bibinfo{year}{2020}.
\newblock \bibinfo{title}{On the use of lorawan in smart cities: A study with blocking interference}.
\newblock \bibinfo{journal}{IEEE Internet of Things Journal} \bibinfo{volume}{7}, \bibinfo{pages}{2806--2815}.
\newblock \DOIprefix\doi{10.1109/JIOT.2019.2962976}.
\bibitem[{MetaGeek(2024)}]{metaGeek}
\bibinfo{author}{MetaGeek, I.}, \bibinfo{year}{2024}.
\newblock \bibinfo{title}{Zigbee and wi-fi coexistence}.
\newblock \URLprefix \url{https://www.metageek.com/training/resources/zigbee-wifi-coexistence/}.
\bibitem[{Oikonomou et~al.(2022)Oikonomou, Duquennoy, Elsts, Eriksson, Tanaka and Tsiftes}]{Contiki-NG-2022}
\bibinfo{author}{Oikonomou, G.}, \bibinfo{author}{Duquennoy, S.}, \bibinfo{author}{Elsts, A.}, \bibinfo{author}{Eriksson, J.}, \bibinfo{author}{Tanaka, Y.}, \bibinfo{author}{Tsiftes, N.}, \bibinfo{year}{2022}.
\newblock \bibinfo{title}{The {Contiki-NG} open source operating system for next generation {IoT} devices}.
\newblock \bibinfo{journal}{SoftwareX} \bibinfo{volume}{18}, \bibinfo{pages}{101089}.
\newblock \DOIprefix\doi{https://doi.org/10.1016/j.softx.2022.101089}.
\bibitem[{Ojeda et~al.(2023)Ojeda, Mendez, Fajardo and Ellinger}]{ojeda_sensor_model_2023}
\bibinfo{author}{Ojeda, F.}, \bibinfo{author}{Mendez, D.}, \bibinfo{author}{Fajardo, A.}, \bibinfo{author}{Ellinger, F.}, \bibinfo{year}{2023}.
\newblock \bibinfo{title}{On wireless sensor network models: A cross-layer systematic review}.
\newblock \bibinfo{journal}{Journal of Sensor and Actuator Networks} \bibinfo{volume}{12}.
\newblock \URLprefix \url{https://www.mdpi.com/2224-2708/12/4/50}, \DOIprefix\doi{10.3390/jsan12040050}.
\bibitem[{Phan et~al.(2019)Phan, Kim, Kim, Lee and Ham}]{phan_performance_tsych_2019}
\bibinfo{author}{Phan, L.A.}, \bibinfo{author}{Kim, T.}, \bibinfo{author}{Kim, T.}, \bibinfo{author}{Lee, J.}, \bibinfo{author}{Ham, J.H.}, \bibinfo{year}{2019}.
\newblock \bibinfo{title}{Performance analysis of time synchronization protocols in wireless sensor networks}.
\newblock \bibinfo{journal}{Sensors} \bibinfo{volume}{19}.
\newblock \URLprefix \url{https://www.mdpi.com/1424-8220/19/13/3020}, \DOIprefix\doi{10.3390/s19133020}.
\bibitem[{Raj(2021)}]{Raj_indoor_rssi_2021}
\bibinfo{author}{Raj, N.}, \bibinfo{year}{2021}.
\newblock \bibinfo{title}{Indoor rssi prediction using machine learning for wireless networks}, in: \bibinfo{booktitle}{2021 International Conference on COMmunication Systems \& NETworkS (COMSNETS)}, pp. \bibinfo{pages}{372--374}.
\newblock \DOIprefix\doi{10.1109/COMSNETS51098.2021.9352852}.
\bibitem[{Reitz et~al.(2024)Reitz, Künzle, Franchi and Lübke}]{Reitz_FMCW_interference_2024}
\bibinfo{author}{Reitz, P.}, \bibinfo{author}{Künzle, C.}, \bibinfo{author}{Franchi, N.}, \bibinfo{author}{Lübke, M.}, \bibinfo{year}{2024}.
\newblock \bibinfo{title}{Evaluation of the interference performance of fmcw radar sensors in dense indoor environments}.
\newblock \bibinfo{journal}{IEEE Access} \bibinfo{volume}{12}, \bibinfo{pages}{46834--46850}.
\newblock \DOIprefix\doi{10.1109/ACCESS.2024.3382547}.
\bibitem[{Romanov et~al.(2020)Romanov, Gringoli and Sikora}]{romanov2020precise}
\bibinfo{author}{Romanov, A.M.}, \bibinfo{author}{Gringoli, F.}, \bibinfo{author}{Sikora, A.}, \bibinfo{year}{2020}.
\newblock \bibinfo{title}{A precise synchronization method for future wireless tsn networks}.
\newblock \bibinfo{journal}{IEEE Transactions on Industrial Informatics} \bibinfo{volume}{17}, \bibinfo{pages}{3682--3692}.
\bibitem[{Salazar-Lopez et~al.(2024)Salazar-Lopez, Millan-Almaraz, Gaxiola-Camacho, Vazquez-Becerra and Leal-Graciano}]{Salazar_Lopez_gps_synchronization_2024}
\bibinfo{author}{Salazar-Lopez, J.R.}, \bibinfo{author}{Millan-Almaraz, J.R.}, \bibinfo{author}{Gaxiola-Camacho, J.R.}, \bibinfo{author}{Vazquez-Becerra, G.E.}, \bibinfo{author}{Leal-Graciano, J.M.}, \bibinfo{year}{2024}.
\newblock \bibinfo{title}{Gps-based network synchronization of wireless sensors for extracting propagation of disturbance on structural systems}.
\newblock \bibinfo{journal}{Sensors} \bibinfo{volume}{24}.
\newblock \URLprefix \url{https://www.mdpi.com/1424-8220/24/1/199}, \DOIprefix\doi{10.3390/s24010199}.
\bibitem[{Shen et~al.(2021)Shen, Liu, Ni, Liu, Zhao and Shang}]{Shen_link_correlation_2021}
\bibinfo{author}{Shen, X.}, \bibinfo{author}{Liu, L.}, \bibinfo{author}{Ni, Z.}, \bibinfo{author}{Liu, M.}, \bibinfo{author}{Zhao, B.}, \bibinfo{author}{Shang, Y.}, \bibinfo{year}{2021}.
\newblock \bibinfo{title}{Link-correlation-aware opportunistic routing in low-duty-cycle wireless networks}.
\newblock \bibinfo{journal}{Sensors} \bibinfo{volume}{21}.
\newblock \URLprefix \url{https://www.mdpi.com/1424-8220/21/11/3840}, \DOIprefix\doi{10.3390/s21113840}.
\bibitem[{Shi and Li(2017a)}]{Shi_FunZigBee_Wifi_2017}
\bibinfo{author}{Shi, G.}, \bibinfo{author}{Li, K.}, \bibinfo{year}{2017}a.
\newblock \bibinfo{title}{Fundamentals of ZigBee and WiFi}. \bibinfo{publisher}{Springer International Publishing}, \bibinfo{address}{Cham}.
\newblock pp. \bibinfo{pages}{9--27}.
\newblock \URLprefix \url{https://doi.org/10.1007/978-3-319-47806-7_2}, \DOIprefix\doi{10.1007/978-3-319-47806-7_2}.
\bibitem[{Shi and Li(2017b)}]{shi2017signal}
\bibinfo{author}{Shi, G.}, \bibinfo{author}{Li, K.}, \bibinfo{year}{2017}b.
\newblock \bibinfo{title}{Signal interference in WiFi and ZigBee networks}.
\newblock \bibinfo{publisher}{Springer}.
\bibitem[{Simon O.~Haykin(2004)}]{Haykin_wireless_communication}
\bibinfo{author}{Simon O.~Haykin, M.M.}, \bibinfo{year}{2004}.
\newblock \bibinfo{title}{Modern Wireless Communications: International Edition}.
\newblock \bibinfo{publisher}{Pearson}.
\bibitem[{Sondej and Bednarczyk(2024)}]{Sondej_ULP_tsynch_2024}
\bibinfo{author}{Sondej, T.}, \bibinfo{author}{Bednarczyk, M.}, \bibinfo{year}{2024}.
\newblock \bibinfo{title}{Ultra-low-power sensor nodes for real-time synchronous and high-accuracy timing wireless data acquisition}.
\newblock \bibinfo{journal}{Sensors} \bibinfo{volume}{24}.
\newblock \URLprefix \url{https://www.mdpi.com/1424-8220/24/15/4871}, \DOIprefix\doi{10.3390/s24154871}.
\bibitem[{SonicWall(2024)}]{sonicwall_rssi}
\bibinfo{author}{SonicWall}, \bibinfo{year}{2024}.
\newblock \bibinfo{title}{Wireless: Snr, rssi and noise basics of wireless troubleshooting}.
\newblock \URLprefix \url{https://www.sonicwall.com/support/knowledge-base/wireless-snr-rssi-and-noise-basics-of-wireless-troubleshooting/180314090744170/}.
\bibitem[{Stancu et~al.(2020)Stancu, Halunga, Fratu, Florea, Berceanu and Cristian}]{Stancu_spetral_analysis_wifi_2020}
\bibinfo{author}{Stancu, E.}, \bibinfo{author}{Halunga, S.}, \bibinfo{author}{Fratu, O.}, \bibinfo{author}{Florea, C.}, \bibinfo{author}{Berceanu, M.G.}, \bibinfo{author}{Cristian, C.}, \bibinfo{year}{2020}.
\newblock \bibinfo{title}{Spectral analysis in the 2.4 ghz wifi band in bucharest}, in: \bibinfo{booktitle}{2020 13th International Conference on Communications (COMM)}, pp. \bibinfo{pages}{435--438}.
\newblock \DOIprefix\doi{10.1109/COMM48946.2020.9142040}.
\bibitem[{Suroso et~al.(2022)Suroso, Cherntanomwong and Sooraksa}]{suroso_rssi_synthesis_2022}
\bibinfo{author}{Suroso, D.J.}, \bibinfo{author}{Cherntanomwong, P.}, \bibinfo{author}{Sooraksa, P.}, \bibinfo{year}{2022}.
\newblock \bibinfo{title}{Deep generative model-based rssi synthesis for indoor localization}, in: \bibinfo{booktitle}{2022 19th International Conference on Electrical Engineering/Electronics, Computer, Telecommunications and Information Technology (ECTI-CON)}, pp. \bibinfo{pages}{1--5}.
\newblock \DOIprefix\doi{10.1109/ECTI-CON54298.2022.9795409}.
\bibitem[{Tan et~al.(2020)Tan, Yang, Pang, Gao, Li and Chen}]{Tan_uav_tsynchronization_2020}
\bibinfo{author}{Tan, Z.}, \bibinfo{author}{Yang, X.}, \bibinfo{author}{Pang, M.}, \bibinfo{author}{Gao, S.}, \bibinfo{author}{Li, M.}, \bibinfo{author}{Chen, P.}, \bibinfo{year}{2020}.
\newblock \bibinfo{title}{Uav-assisted low-consumption time synchronization utilizing cross-technology communication}.
\newblock \bibinfo{journal}{Sensors} \bibinfo{volume}{20}.
\newblock \URLprefix \url{https://www.mdpi.com/1424-8220/20/18/5134}, \DOIprefix\doi{10.3390/s20185134}.
\bibitem[{Tinka et~al.(2010)Tinka, Watteyne and Pister}]{tinka2010decentralized}
\bibinfo{author}{Tinka, A.}, \bibinfo{author}{Watteyne, T.}, \bibinfo{author}{Pister, K.}, \bibinfo{year}{2010}.
\newblock \bibinfo{title}{A decentralized scheduling algorithm for time synchronized channel hopping}, in: \bibinfo{booktitle}{Ad Hoc Networks: Second International Conference, ADHOCNETS 2010, Victoria, BC, Canada, August 18-20, 2010, Revised Selected Papers 2}, \bibinfo{organization}{Springer}. pp. \bibinfo{pages}{201--216}.
\bibitem[{Wu et~al.(2011)Wu, Chaudhari and Serpedin}]{wu_clock_synch}
\bibinfo{author}{Wu, Y.C.}, \bibinfo{author}{Chaudhari, Q.}, \bibinfo{author}{Serpedin, E.}, \bibinfo{year}{2011}.
\newblock \bibinfo{title}{Clock synchronization of wireless sensor networks}.
\newblock \bibinfo{journal}{IEEE Signal Processing Magazine} \bibinfo{volume}{28}, \bibinfo{pages}{124--138}.
\newblock \DOIprefix\doi{10.1109/MSP.2010.938757}.
\bibitem[{Yang et~al.(2016)Yang, Yan, Li, Zhang, Tao and You}]{Yang_cti_wifi_zigbee}
\bibinfo{author}{Yang, P.}, \bibinfo{author}{Yan, Y.}, \bibinfo{author}{Li, X.Y.}, \bibinfo{author}{Zhang, Y.}, \bibinfo{author}{Tao, Y.}, \bibinfo{author}{You, L.}, \bibinfo{year}{2016}.
\newblock \bibinfo{title}{Taming cross-technology interference for wi-fi and zigbee coexistence networks}.
\newblock \bibinfo{journal}{IEEE Transactions on Mobile Computing} \bibinfo{volume}{15}, \bibinfo{pages}{1009--1021}.
\newblock \DOIprefix\doi{10.1109/TMC.2015.2442252}.
\bibitem[{Yin et~al.(2018)Yin, Li, Kim and He}]{yin2018explicit}
\bibinfo{author}{Yin, Z.}, \bibinfo{author}{Li, Z.}, \bibinfo{author}{Kim, S.M.}, \bibinfo{author}{He, T.}, \bibinfo{year}{2018}.
\newblock \bibinfo{title}{Explicit channel coordination via cross-technology communication}, in: \bibinfo{booktitle}{Proceedings of the 16th Annual International Conference on Mobile Systems, Applications, and Services}, pp. \bibinfo{pages}{178--190}.
\bibitem[{Zacharias et~al.(2012)Zacharias, Newe, O’Keeffe and Lewis}]{Zacharias_rssi_traces_2012}
\bibinfo{author}{Zacharias, S.}, \bibinfo{author}{Newe, T.}, \bibinfo{author}{O’Keeffe, S.}, \bibinfo{author}{Lewis, E.}, \bibinfo{year}{2012}.
\newblock \bibinfo{title}{Identifying sources of interference in rssi traces of a single ieee 802.15.4 channel}.
\bibitem[{Zhang et~al.(2020)Zhang, Yuan and Bi}]{Zhang_design_tsynchronization_2020}
\bibinfo{author}{Zhang, D.}, \bibinfo{author}{Yuan, Y.}, \bibinfo{author}{Bi, Y.}, \bibinfo{year}{2020}.
\newblock \bibinfo{title}{A design of a time synchronization protocol based on dynamic route and forwarding certification}.
\newblock \bibinfo{journal}{Sensors} \bibinfo{volume}{20}.
\newblock \URLprefix \url{https://www.mdpi.com/1424-8220/20/18/5061}, \DOIprefix\doi{10.3390/s20185061}.
\bibitem[{Zhu et~al.(2010)Zhu, Zhong, He and Zhang}]{Ting_link_correlation_2010}
\bibinfo{author}{Zhu, T.}, \bibinfo{author}{Zhong, Z.}, \bibinfo{author}{He, T.}, \bibinfo{author}{Zhang, Z.L.}, \bibinfo{year}{2010}.
\newblock \bibinfo{title}{Exploring link correlation for efficient flooding in wireless sensor networks}, in: \bibinfo{booktitle}{7th USENIX Symposium on Networked Systems Design and Implementation (NSDI 10)}, \bibinfo{publisher}{USENIX Association}, \bibinfo{address}{San Jose, CA}.
\newblock \URLprefix \url{https://www.usenix.org/conference/nsdi10-0/exploring-link-correlation-efficient-flooding-wireless-sensor-networks}.
\bibitem[{Zuo et~al.(2021)Zuo, Xie, Zhang and Yang}]{Zuo_uav_wifi_recognition_2021}
\bibinfo{author}{Zuo, M.}, \bibinfo{author}{Xie, S.}, \bibinfo{author}{Zhang, X.}, \bibinfo{author}{Yang, M.}, \bibinfo{year}{2021}.
\newblock \bibinfo{title}{Recognition of uav video signal using rf fingerprints in the presence of wifi interference}.
\newblock \bibinfo{journal}{IEEE Access} \bibinfo{volume}{9}, \bibinfo{pages}{88844--88851}.
\newblock \DOIprefix\doi{10.1109/ACCESS.2021.3089590}.

\end{thebibliography}
 \nocite{*}

\end{document}